\documentstyle[12pt]{article}

\makeatletter

\def\section{\@startsection {section}{1}{\z@}{-3.5ex plus -1ex minus
    -.2ex}{5ex plus .2ex}{\Large\centering\bf}}
\def\subsection{\@startsection{subsection}{2}{\z@}{-3.25ex plus -1ex minus
   -.2ex}{3ex plus .2ex}{\large\centering\bf}}
\def\subsubsection{\@startsection{subsubsection}{3}{\z@}{-3.25ex plus
 -1ex minus -.2ex}{1.5ex plus .2ex}{\normalsize\centering\bf}}
\def\paragraph{\@startsection
     {paragraph}{4}{\z@}{3.25ex plus 1ex minus .2ex}{-1em}{\normalsize\bf}}
\def\subparagraph{\@startsection
     {subparagraph}{4}{\parindent}{3.25ex plus 1ex minus
     .2ex}{-1em}{\normalsize\em}}

\def\maketitle{\par
 \begingroup
   \def\thefootnote{\fnsymbol{footnote}}
   \def\@makefnmark{\hbox
       to 0pt{$^{\@thefnmark}$\hss}}
   \if@twocolumn
     \twocolumn[\@maketitle]
     \else \newpage
     \global\@topnum\z@        % Prevents figures from going at top of page.
     \@maketitle \fi\thispagestyle{plain}\@thanks
 \endgroup
 \setcounter{footnote}{0}
 \let\maketitle\relax
 \let\@maketitle\relax
 \gdef\@thanks{}\gdef\@author{}\gdef\@title{}\let\thanks\relax}

\def\@maketitle{\newpage
 \null
 \vskip 2em                 % Vertical space above title.
 \begin{center}
  {\LARGE \@title \par}     % Title set in \LARGE size.       %%%
  \vskip 1.5em                % Vertical space after title.
  {\large                       % each author set in \large, in a   %%%
   \lineskip .5em           % tabular environment
   \begin{tabular}[t]{c}\@author
   \end{tabular}\par}
  \vskip 1em              % Vertical space after author.
  {\large \@date}           % Date set in \large size.    %%%
\end{center}
 \par
 \vskip 1.5em}                % Vertical space after date.

\def\abstract{\if@twocolumn
\section*{Abstract}
\else \small
\begin{center}
{\bf Abstract\vspace{-.5em}\vspace{0pt}}
\end{center}
\quotation
\fi}

\def\endabstract{\if@twocolumn\else\endquotation\fi}

 \newlength{\captionwidthh}

 \long\def\@caption#1[#2]#3{\addcontentsline{\csname
  ext@#1\endcsname}{#1}{\protect\numberline{\csname
  the#1\endcsname}{\ignorespaces #2}}\par
    \begingroup
      \setlength{\captionwidthh}{\textwidth}
      \addtolength{\captionwidthh}{-2cm}
      \begin{center}
      \parbox{\captionwidthh}{
	\small
        \@makecaption{\csname fnum@#1\endcsname}{\ignorespaces #3}\par
      }
      \end{center}

    \endgroup }

\def\@cite#1{#1}            	% Produces the output of the \cite command.

\def\thebibliography#1{\section*{References\@mkboth
  {REFERENCES}{REFERENCES}}\list
  {[\arabic{enumi}]}{\settowidth\labelwidth{#1}\leftmargin\labelwidth
    \advance\leftmargin\labelsep
    \addtolength\itemindent{-\labelwidth}
    \usecounter{enumi}}
    \def\newblock{\hskip .11em plus .33em minus .07em}
    \sloppy\clubpenalty4000\widowpenalty4000
    \sfcode`\.=1000\relax}

\def\@lbibitem[#1]#2{\item[]\if@filesw
      { \def\protect##1{\string ##1\space}\immediate
        \write\@auxout{\string\bibcite{#2}{#1}}}\fi\ignorespaces}
\makeatother

\newcommand{\la}[1]{\label{#1}}

\newcommand{\pspict}[2]{
  \vspace*{#1}
  \special{dvitops: import #2 \the\textwidth #1}
}

\newcommand{\zerotxt}[1]{
	\mbox{$\lefteqn{\mbox{#1}}$}
}

\newlength{\numpit}

\newlength{\indeksinpituus}

\newlength{\ypit}

\newcommand{\be}{\begin{equation}}
\newcommand{\ee}{\end{equation}}
\newcommand{\ba}{\begin{eqnarray}}
\newcommand{\ea}{\end{eqnarray}}

\newcommand{\etal}{{\em et al.\/}\ }
\newcommand{\eq}{eq.~}
\newcommand{\eqs}{eqs.~}
\newcommand{\fig}{fig.~}

\newcommand{\dd}{{\rm d}}

\newlength{\pohjaviiva}
\newcommand{\phibar}{\bar{\phi}}
\newcommand{\lbar}{\bar{\lambda}}
\newcommand{\vsas}{\vspace*{0.0cm}}
\newcommand{\vsass}{\vspace*{1cm}}

\newcommand{\confps}[1]{
\vspace{7.5cm}
\special{dvitops: import #1 \the\textwidth 10cm}
\vspace{-3.5cm}
}

\begin{document}

%%% K-A's titlepage with *lots* of my changes %%%%%%%%%%%%%%%%%%%%%%%%%%%%%%%

%\newfont{\densebold}{cmb10} - doesn't have any effect
%\newfont{\pstimesroman}{ptmb} - produces an error announcement

%\setlength{\topmargin}{-10mm}
%\setlength{\headheight}{5mm}
%\setlength{\headsep}{15mm}
%\setlength{\textheight}{230mm}

%\setlength{\oddsidemargin}{-0.5mm}
%\setlength{\evensidemargin}{-0.5mm}
%\setlength{\textwidth}{160mm}
%\setlength{\footskip}{15mm}
%\setlength{\abovedisplayskip}{20mm}
%\setlength{\abovedisplayshortskip}{20mm}
%\setlength{\belowdisplayskip}{20mm}
%\setlength{\belowdisplayshortskip}{20mm}
%\setlength{\parsep}{0mm}
%\setlength{\listparindent}{0mm}

%\setlength{\baselineskip}{9mm}
%\renewcommand{\baselinestretch}{1.25}

\pagestyle{empty}

%\begin{document}

%\begin{titlepage}

\setcounter{page}{-1}

%\vspace*{1mm}
\vspace*{-12mm}

\noindent\parbox[t]{61mm}
{UNIVERSITY OF HELSINKI \mbox{}}
\  \hfill  \
\begin{minipage}[t]{55mm}
\begin{flushleft}
REPORT
\vspace*{-2.5mm}
SERIES IN
%\\
   THEORETICAL PHYSICS
\end{flushleft} \end{minipage}

\vspace*{9mm}

\begin{center}

\noindent HU-TFT-IR-93-1

%\vspace*{30mm}
\vspace*{45mm}

%\vfill

{\huge\bf COSMOLOGICAL}

\vspace*{2mm}
{\huge\bf PHASE~TRANSITIONS}
\vspace*{4mm}

\vspace*{24mm}

\noindent{\Large\bf JANNE~~IGNATIUS}

\vspace*{8mm}

{\large\bf Department of Theoretical Physics}

\vspace*{-2mm} {\large\bf Faculty of Science}

\vspace*{-2mm} {\large\bf University of Helsinki}

\vspace*{-2mm} {\large\bf Helsinki, Finland}

\vspace*{22mm}

%\vfill

%{\large\rm\noindent Academic dissertation}
%\bigskip

{\small\it ACADEMIC DISSERTATION}
\smallskip

\vspace*{-2mm}
{\small\it
To be presented, with the permission of the Faculty of Science}

\vspace*{-2mm}
{\small\it of the University of Helsinki, for public criticism
in Auditorium XII}

\vspace*{-2mm}
{\small\it on October 30th, 1993, at 10 o'clock.}

%\vspace*{20mm}
%\vfill
\vspace*{24mm}

{\large\noindent Helsinki 1993}

%\vspace*{20mm}

\end{center}

\normalsize\rm

%\end{titlepage}
\newpage

%\setlength{\parskip}{2mm}

%\pagenumbering{roman}

%%% Borrowed from Kodis's titlepage - not used %%%%%%%%%%%%%%%%%%%%%%%%%%%%%%%

%\begin{titlepage}
%\centering

%\vspace*{0.5cm}
%%Commentationes Physico-Mathematicae XXX/1990\\
%%Dissertationes No XX\\
%%\vspace*{4cm}
%\vspace*{3cm}

%{\large
%COSMOLOGICAL\ \ PHASE\ \ TRANSITIONS\\
%}

%%\vspace{2cm}
%\vspace{3cm}
%{\large Janne Ignatius}

%\vspace{1cm}
%{\em
%Department of Theoretical Physics\\
%Faculty of Science\\
%University of Helsinki\\
%Helsinki, Finland
%}

%\vspace{2.5cm}
%Academic dissertation\\
%{\small \em
%To be presented, with the permission of the Faculty of Science\\
%of the University of Helsinki, for public criticism in Auditorium ??\\
%on Month? ??th, 199?, at 12 o'clock noon.
%}

%\vspace{0.5cm}
%Helsinki 199?

%\end{titlepage}

%%% Normal part begins (format mainly borrowed from Kodis)%%%%%%%%%%%%%%%%%%%%

\vspace*{20cm}

\begin{center}

\vspace*{-3.5mm}
{\footnotesize\bf
 ISBN 951-45-6526-6}

\vspace*{-3.5mm}
{\footnotesize\bf
 ISSN 0786-2547}

\vspace*{-3.5mm}
{\footnotesize\bf
 Helsinki 1993}

\vspace*{-3.5mm}
{\footnotesize\bf
 University Press}

\end{center}

\newpage

\pagestyle{plain}

\setlength{\parskip}{0mm}

%\vspace*{-1.5cm}

\section*{Preface}
	\addtocontents{toc}{\protect\addvspace{3mm}}
	\addcontentsline{toc}{section}{Preface}
	\addtocontents{toc}{\protect\addvspace{3mm}}

\vspace*{-0.5cm}

This Thesis is based on research done
%at the University of Helsinki,
at the Department of Theoretical Physics,
%at University of Helsinki,
and since August 1993, at the Research Institute for Theoretical
Physics, University of Helsinki.
While at the Department, I have been funded
by the University, the Academy of Finland, the Finnish Cultural Foundation and
the Magnus Ehrnrooth Foundation. I wish to express my gratitude to
these institutes and foundations, and to ITP
at Santa Barbara for an inspiring visit in April--May 1992.

%It is with great pleasure that I take the opportunity to express my deep
%gratitude to
First and foremost, I want to thank my advisor and collaborator
Professor Keijo Kajantie. His excellent guidance has been invaluable to me.

I thank Docent Kari Rummukainen for enlightening discussions
and correspondence---and for tutoring me
%as a freshman
in the beginning of my studies at the University.
Likewise, thanks are due to Professor Kari Enqvist for good co-authorship.

I would like to warmly thank Mr.\ Mikko Laine for conversations and careful
reading  of the manuscript, and Doc.\ Hannu Kurki-Suonio for discussions.
I am deeply indebted to Doc.\ Jukka Maalampi and Prof.\ Vesa Ruuskanen
for thorough inspection of the manuscript and numerous useful comments.

%Likewise, warm thanks are due to ...
%... for their kind support and inspiring co-authorship
%... valuable discussions and guidance

I wish to thank Professors Christofer Cronstr\"om and Paul Hoyer for their
support and interest on my career, and Prof.\ Juhani Keinonen for guidance
in the course of my first research project.
I am grateful to Doc.\ Tapio Ala-Nissil\"a,
Doc.\ Kimmo Kainulainen, Dr.\ Leo K\"arkk\"ainen, Doc.\ Claus Montonen,
Doc.\ Erkki Pajanne, Doc.\ Martti Salomaa and Dr.\ Kalle-Antti Suominen
for discussions and help.

I thank my colleagues and the personnel at
%friends at the University
the various physics institutes for providing the facilities  and
the unique atmosphere,
all my friends at the University
for making these first seven years of my academic life
%a pleasant and rewarding
an enjoyable
period.

My special thanks
%are due
go
to my parents Heikki and Ann
%for offering a home in which it was natural to grow into science,
from whom I have learned to wonder
%the secrets of Nature,
the nature around us,
and to my friend Eve.
%for ...
\medskip

This Thesis is dedicated to all thinking creatures in our Universe.

\vspace{0.5cm}

%\noindent
%Helsinki, September 1993
%
%\noindent
%{\em Janne Ignatius}

\noindent\hspace*{94mm}\begin{minipage}[t]{49mm}Helsinki, 3 October 1993\\
{\it Janne Ignatius}\end{minipage}

\newpage

\setlength{\parskip}{2mm}

%% Abstract
\setlength{\pohjaviiva}{\baselineskip}
\setlength{\baselineskip}{6mm}
\noindent
{\bf Cosmological Phase Transitions }\\
Jan Jalmar (\underline{Janne}) Ignatius \\
University of Helsinki, 1993

\vspace{0.5cm}
\vspace*{-0.6cm}
\section*{Abstract}
	\addcontentsline{toc}{section}{Abstract}
	\addtocontents{toc}{\protect\addvspace{3mm}}

\vspace*{-0.5cm}

It is generally believed that several phase transitions have taken place
in the early Universe.
The effects of cosmological phase transitions may well have been
crucial for the evolution of the Universe, and thus for the existence of life
as we know it.

The cosmological phase transitions investigated here are related to strong and
electroweak interactions.
When the Universe was about $10^{-11}$ seconds old and
the horizon radius equaled one centimeter,
symmetry between weak and electromagnetic interactions was broken.
It is quite possible that the baryon asymmetry, one of the most
important properties of our Universe, was generated in this transition.
Later happened the phase transition from the quark--gluon plasma to the hadron
matter. At the quark--hadron transition the size of a causally connected
region of space, the horizon radius, was ten kilometers and the age
of the Universe $10^{-5}$ seconds.

The new Z(3) phase transition suggested in this work is situated to
a temperature or energy two orders of magnitude above the electroweak scale.
At that time the Universe was roughly $10^{-15}$ seconds old and the horizon
radius was of the order of one micrometer.
This hypothetical phase transition is caused purely by
the high-temperature properties of strong interactions.

The discussion of the different phase transitions is based on the assumption
that they are first order.
It is pointed out that especially the onset of a cosmological
phase transition shows a universal behavior.
%allowing for a general approach.
General methods are presented, applicable to an analysis of the sequence
of events taking place in cosmological first-order phase transitions.

An equation of state is derived for the electroweak matter near the
phase transition point.
The thermodynamically allowed region for the  velocities
of the phase transition front is determined.

The nucleation rate of bubbles of the broken-symmetry
phase is computed in generic first-order cosmological phase transitions.
The radial dependence of the Helmholtz free energy of the bubbles is also
discussed.

\clearpage
\setlength{\baselineskip}{\pohjaviiva}

\tableofcontents
\newpage

%% List of publications

\section*{List of Papers}
	\addcontentsline{toc}{section}{List of Papers}
	\addtocontents{toc}{\protect\addvspace{3mm}}

This Thesis consists of an introductory review part,
followed by three research publications in chronological order:
%; they are referred with roman numerals:
\medskip

\begin{list}{}{\setlength{\labelsep}{13mm} \setlength{\leftmargin}{15mm}
                 \setlength{\labelwidth}{15mm}}

\item[\zerotxt{I:}] K.~Enqvist, J.~Ignatius, K.~Kajantie and K.~Rummukainen, \\
{\em Nucleation and Bubble Growth in a First-Order Cosmological Electroweak
Phase Transition}, \\ Physical Review D {\bf 45} (1992) 3415--3428. \bigskip

\item[\zerotxt{II:}]
J.~Ignatius, K.~Kajantie and K.~Rummukainen, \\
{\em Cosmological QCD Z(3) Phase Transition in the 10 TeV Temperature
Range?}, \\
Physical Review Letters {\bf 68} (1992) 737--740.
\bigskip

\item[\zerotxt{III:}]
J.~Ignatius, \\
{\em Bubble Free Energy in Cosmological Phase Transitions}, \\
Physics Letters B {\bf 309} (1993) 252--257.
\bigskip

\end{list}

\clearpage

%\setcounter{page}{1}
%\pagenumbering{arabic}

%% Introduction

\vsas
\section{Introduction}   \label{1}

Cosmological phase transitions offer a rich variety of physical
phenomena for investigation, and
some of their effects may be observable in the present Universe.
In this Thesis the mechanisms of cosmological phase transitions
are studied, concentrating on transitions which are related to strong
and electroweak interactions.
%and discuss their cosmological consequences.
%The subjects of study of this Thesis are the cosmological phase transitions
%related to strong and electroweak interactions.
%This Thesis considers mechanisms of cosmological phase transitions
%related to strong and electroweak interactions.

All the phase transitions to be discussed are---or
are assumed to be---first order.
In the first-order transitions a metastable phase may exist alongside
a stable one for some temperature range.
In the thermodynamical limit, {\em i.e.}, in an infinitely large
system in which the temperature is changed with an infinitely slow rate,
the phase transition takes place at a certain critical temperature~$T_c$.

If the rate of the temperature decrease is
finite, as is the case in the expanding Universe,
the phase transition temperature differs from the equilibrium
value~$T_c$.
%As the Universe grows older, it cools down.
At the critical temperature
of a phase transition nothing happens, the high-temperature phase
just moves into a supercooled state.
At a somewhat lower temperature bubbles of the new
phase begin to nucleate. The bubbles grow and convert the space to the new
phase.
%At any temperature below $T_c$,
The new phase has a lower energy
density than the old phase. This means that in the phase
transition the Universe is heated up to a certain temperature not higher
than~$T_c$. After the transition is completed, the Universe starts
to cool again in the usual way.

The supercooling is crucial for the scenarios in which
the baryon asymmetry of the Universe is generated at the electroweak scale.
%Baryon asymmetry means that
There is more matter than antimatter in the
Universe, and more than a billion photons for every baryon.
This fundamental cosmological fact,
%quite essential
crucial
%for our Universe and
for us human beings, should be explained in a satisfactory way.

A few years ago it was realized that at temperatures above the critical
temperature of the electroweak theory, certain
electroweak processes mediated by the so-called sphalerons destroy any
pre-existing baryon plus lepton number asymmetry
[\cite{KuzminRubakovShaposhnikov85}].
It became thus necessary to understand how the baryon asymmetry of the Universe
could have been created at the electroweak phase transition.
In principle, this is possible since all the three necessary conditions
for generation of baryon asymmetry [\cite{Sakharov67}]
could have been satisfied during the transition.
Firstly, due to the anomaly in the electroweak theory
[\cite{tHooft76a}, 1976b],
baryon-number violating reactions were taking place.
Secondly, CP--symmetry was violated because of
%the electroweak interactions.
fundamental gauge and Higgs interactions of quarks.
The third condition,
%requirement of
a departure from thermal
equilibrium, was well satisfied because of the supercooling,
provided the phase transition is first order.

In order to obtain any quantitative estimates
for the amount of baryon asymmetry created, one must have a detailed
understanding of how the electroweak phase transition proceeded.
%This has been the motivation for a substantial portion of the research
%presented in this Thesis.
Motivated by this, we have investigated in this Thesis mechanisms of the
electroweak phase transition.

The other phase transitions considered in this work are related
to quantum chromodynamics.
The possible observable consequences of the cosmological quark--hadron phase
transition are due to the density inhomogeneities produced
during the transition.
If the length scale of these inhomogeneities had been large enough,
they could have later affected the nucleosynthesis. This effect could
be observed in the abundance of light elements in the present-day Universe.
%Although not completely excluded, this does not seem to be the case,
However, it seems probable that the length scale was too small for that,
as will be discussed later on.
%(Section~\ref{4.2}).
In the case of the
Z(3) phase transition suggested in this Thesis, it might
in principle be that the density inhomogeneities generated could have affected
later processes like the electroweak phase transition.

%The purpose of this Thesis
%is by no means to give an overview of all the topics
%related to cosmological phase transitions.
%We do not attempt to present any overview of all the topics related to
%cosmological phase transitions in this Thesis.
In addition to those questions considered in this work there are
several other interesting topics related to cosmological phase transitions.
For example, we have not studied the
possible phase transition of a grand unified theory, which may have been
cosmologically important as a driving source of inflation.
Likewise, the topological defects, like
monopoles or cosmic strings, which could have been created in
cosmological first-order phase transitions are not discussed.

This Thesis is organized as follows.
We first give a brief summary of the contents of
%In the remainder of this Section, an introduction to the
the original research papers, which are appended.
In Section~\ref{2} the main events in
the evolution of the Universe are described. In Section~\ref{3} cosmological
first-order phase transitions are discussed on a general level,
without specifying the physical model.
%First the field-theoretic expression for the
%nucleation rate is introduced. Then the nucleation and growth of bubbles are
%discussed, and equations determining the degree of supercooling and
%the average distance between nucleation sites are given. Finally, the
%preceding results are rewritten in the limit of small relative supercooling.
%Several comments related to Papers~I--III are also made in that Section.
In Section~\ref{4} the general methods presented in the previous Section
are applied to
two physical cases, to the electroweak and the quark--hadron phase transition.
%The concluding words appear in Section~\ref{5},
%followed by the research papers.
Finally, in Section~\ref{5} we present conclusions and point out directions for
the future work.
%The original research publications are appended in the end.

\bigskip

\vsass
\subsection*{Summary of the Original Papers}   \label{1.1}
	\addcontentsline{toc}{subsection}{Summary of the Original Papers}
	\addtocontents{toc}{\protect\addvspace{3mm}}

\medskip

\paragraph{Paper~I:
Nucleation and Bubble Growth in a First-Order Cosmological
Electroweak Phase Transition.}
In this paper the thermodynamical properties of electroweak matter
near the critical temperature are systematically investigated
for the first time.
%for the first time in the literature.
Assuming a quartic form for the Higgs
potential (to be discussed in Subsection~\ref{4.1} of this introductory review
part), we derive an equation of
state that describes the electroweak phase transition,
and compare the electroweak transition with the quark--hadron transition.
The nucleation rate of bubbles of
the broken-symmetry phase is computed by solving numerically the field
equation. We present a useful expression for the volume fraction
not touched by the bubbles, slightly different from those given previously by
other authors. We perform
numerical simulations of bubble nucleation and growth which confirm our
analytical calculations.
Finally, we also study what velocities of the phase front
are allowed assuming only that the general conditions of energy-momentum
conservation and entropy increase are valid.
Based on these considerations, we claim
that in the cosmological electroweak phase transition the bubbles most likely
grew as weak deflagrations.
\medskip

\paragraph{Paper~II:
Cosmological QCD Z(3) Phase Transition in the 10 TeV Temperature Range?}
Using as a starting point the earlier observation that in the QCD
%with the quarks included
there are metastable vacua at high temperatures,
we develop a cosmological scenario which leads
to a phase transition, not known before, at a temperature
%of the order of 10~TeV
two orders of magnitude above the electroweak scale.
Qualitatively, this phase transition
%(supposing it took place)
differs from the usual ones in that the pressure difference
between the stable and metastable vacua is huge,
and in that there were only relatively few bubbles
nucleated inside the horizon.
Our scenario is based on the hypothesis that at very early times
domains of metastable vacua were created and underwent an inflationary
expansion
%exponentially,
due to some processes which could be related
for instance to the breaking of the grand unified symmetry.
This hypothesis is the main uncertainty in our scenario.
Later on it has been also claimed that the metastable vacua should
only be interpreted as field configurations that contribute to
the Euclidean path integral, not as physically accessible states
[\cite{Belyaevetal92}; \cite{ChenDobroliubovSemenoff92}].
In a more recent investigation it has been, however,
argued that the metastable vacua do
represent physically realizable systems [\cite{GockschPisarski93}].
\medskip

\paragraph{Paper~III:
Bubble Free Energy in Cosmological Phase Transitions.}
In this paper the free energy of spherical bubbles is studied in order
parameter or Higgs field models having the same quartic potential
as used in Paper~I\@.
A numerical function with a good accuracy is given for the nucleation action.
Using this nucleation free energy of critical bubbles
as an input, the general
free energy is solved as a function of the bubble radius and the temperature.
%In other words, in the first place one knows only the height of
%the free energy wall for forming bubbles of the new phase.
%But applying the method presented in the paper, one is able to determine
%also the radial dependence of the free energy (the profile of the wall).
The calculation is based on the approximation that all the
temperature dependence in the free energy comes from the volume term.
This approximation should be valid if one is not too far from
the limit of small relative supercooling.
%In the paper, the quantities needed for determining
%the general free energy are studied in detail:
The bubble radius and curvature-dependent interface tension
are discussed in detail.
The results of this study are applicable for the case of
the electroweak phase transition, and probably for the
quark--hadron transition as well.

\newpage

\vsas
\section{Short History of the Universe}     \label{2}

The big bang model provides a general framework for describing
the evolution of the Universe. The success of the model is based
on only a few---but fundamental---astronomical observations:
the redshift in the spectra of distant galaxies, the existence of the
2.7 Kelvin cosmic microwave background radiation, and
the abundance of light elements in the Universe.
The redshift is believed to be caused by the expansion of the Universe.
Additional evidence for the big bang model
was presented in April 1992, when it was announced that the observations
from the COBE satellite show an anisotropy of
$\Delta T /T \approx  6 \times 10^{-6}$
in the background radiation [\cite{Smootetal92}].

In principle, we are able to obtain (semi)direct information of the early
Universe by observing electromagnetic or gravitational radiation,
or exotic relics like very massive particles or small black holes.
So far, the only cosmological messengers we have been able to
detect in our instruments are the photons.
This means that in direct observations we
have to limit ourselves to a Universe older than a few hundred
thousand years.
At that time the temperature was somewhat less than the binding energies
of electrons in light atoms. The electrons and light nuclei
were able to form stable atoms, and the Universe became
transparent for photons. This is the epoch we are looking at
when observing the cosmic microwave background radiation.

The abundance of the light elements, as observed in the present-day Universe,
gives us indirect but firm evidence of the time of the primordial
nucleosynthesis.
The major part of the nucleosynthesis
took place when the Universe was a few minutes old
%from $t \simeq 0.01$ to $100$ seconds
[see {\em e.g.}\ \cite{ApplegateHoganScherrer87}].
This is the earliest epoch of which we have more or less certain information.

The study of the very young Universe requires
an extrapolation of the cosmological model to even earlier times.
%times beyond those of which we have good observational information.
In addition to Einstein's theory of gravitation, what is needed
for that is a knowledge of interactions between elementary particles at high
energies. This knowledge, the standard model of particle physics, is based
on laboratory experiments done at
%colossal and very expensive
%very large
colossal particle accelerators.

It is believed that when the Universe was approximately $10^{-5}$ seconds old
and temperature was of the order of $100$~MeV, a phase transition
from the quark--gluon plasma to the hadron matter took place.
In the quark--gluon plasma phase the quarks and gluons were free, whereas
in the hadron phase they became confined and formed mesons like the pions,
and baryons like the proton and the neutron.
The existence of quark--gluon plasma has not yet been confirmed experimentally,
but several groups are trying to detect it by using heavy-ion collisions.
As mentioned in the Introduction, probably the scale of the density
inhomogeneities produced
in the cosmological quark--hadron phase transition was too short
to have observable consequences (see Subsection~\ref{4.2}).
%The cosmological quark--hadron phase transition, supposing it was first order,
%produced density inhomogeneities. However, it seems probable that
%the scale of these density inhomogeneities was too small to have an
%observable effect on the nucleosynthesis.

\begin{figure}[t]
%\pspict{5.5cm}{disk$b:<ignatius.tex.phd>Universe_evol.ps}
\vspace*{5.5cm}
%Sohvi:Die Hard 40/Janne/Kuvia/PhD/Universe_evol
%\vspace*{6cm}
%\special{dvitops: import disk$b:<ignatius.tex.phd>Universe_evol.ps 12cm 6cm}
\caption[a]{Evolution of the Universe.\la{fig:univevol}}
\end{figure}

%The temperature scale of the electroweak phase transition,
%$T \simeq 100$~GeV, is roughly equal to
Further back in time, the electroweak phase transition took place
at $t \simeq 10^{-11}$~s. The temperature was then about $100$~GeV
which roughly corresponds the highest energies available
in particle accelerators. Hence this transition presents
the earliest time in the history of the Universe of which we have
a moderately detailed knowledge.
%some information tested in laboratory experiments.
%In the high-temperature symmetric phase quarks and the
%weak gauge bosons did not have any masses at the classical level
%of the theory.
At the classical level of the electroweak theory, the mass terms of the quarks
and gauge bosons vanish in the high-temperature symmetric phase.
%By naive we mean that the finite-temperature corrections are not taken
%into account.
In the phase transition
%when the symmetry is spontaneously broken by the Higgs mechanism,
their mass terms, which are proportional to
the vacuum expectation value of the Higgs field, become non-vanishing.
%even at the classical level.
It is believed that the electroweak phase transition had a significant
effect on the baryon asymmetry observed in the present Universe.

Of still earlier times we presumably know at least
%the general behavior of the thermodynamical background.
%for example
how temperature evolved during most of the time.
Extrapolating the theories of strong and electroweak interactions
up over ten decades in energy, one is tempted to believe in grand unification
at the scale of $10^{14}$~GeV. However, the huge energy gap between
the electroweak and the grand unified scales
may very well hide new phenomena.
One of the most speculated issues is supersymmetry;
if it exists, the
%models of which predict that the assumed
symmetry between bosons and fermions was restored above certain temperature.
The hypothetical transition suggested in Paper~II is also situated
to the gap between the electroweak and grand unified scales;
its transition temperature
%of about 10~TeV
is two orders of magnitude
above that of the electroweak phase transition.
%in temperature two orders of magnitude above the scale of the electroweak
%transition.
%In the logarithmic axis of \fig\ref{fig:univevol} the transition
%temperature of around 10~TeV is not very far above the electroweak scale.

The most important idea, connected to the (hot) big bang model and
related to these very early times,
is the hypothesis of inflation.
The inflation, or an exponential expansion of the scale factor
at very early times, would resolve the smoothness
problem---why the temperature of the microwave background
is almost uniform over scales bigger than the horizon scale when the photons
last scattered.
Furthermore, the inflation predicts that the
Universe expands eternally but with an always slowing rate,
the alternative we are astonishingly close to according to observations
of the average mass density of the present Universe.

The extreme limit for the validity of the standard cosmology is the
Planck scale.
%For any particle (or a tiny volume of cosmic fluid)
%we can define the Schwarzschild radius which tells the size of the black hole
%associated with that particle. Usually the Schwarzschild radius is much
%smaller than any of the other scales involved, and the ``black hole''
%does not have any relevance. However, if the mass of the particle exceeds
%the Planck mass, the quantum-mechanical wave packet belonging to
%the particle becomes trapped inside the Schwarzschild radius.
Quantum mechanics tells that if we inspect any system on a scale small enough,
the physical quantities fluctuate strongly.
For gravity this fluctuation length scale is given by $1/M_{\rm Pl}$, where
the value of the Planck mass is
$M_{\rm Pl} \approx 1.2 \times 10^{19}$~GeV.\footnote{
 In this Thesis the system of so-called natural units
 is used: $\hbar \!=\! c \!=\! k_B \!=\! 1$.
 However, numerical values for time and length are often given in SI--units.}
The horizon radius, the size of a causally connected region of space,
was equal to this fluctuation scale at the Planck time
when the age of the Universe was $10^{-44}$~s.
At the Planck scale general relativity is not valid any more,
one would need a quantized theory of gravity.

Let us finally present the fundamental equations governing the evolution
of the Universe in the big bang model.
A more detailed treatment can be found for instance in the textbook by
Kolb and Turner~[1990].
A beautiful way to formulate
the Einstein equations is to use the action
\be
  S = \int \dd^4 x \sqrt{-g} \left\{ {\cal L}_{A,\phi,\psi}
     - \frac{1}{16 \pi G} R_{\rm sc}  \right\}   \; , \la{SEH}
\ee
where $g$ is the determinant of the metric tensor,
${\cal L}_{A,\phi,\psi}$ is the Lagrangian density for all the gauge and
matter fields in the standard model,
$G$ is the gravitational constant and $R_{\rm sc}$ the curvature scalar.
The Einstein equations now follow by demanding that the variation of the action
with respect to the metric vanishes.
The metric that produces homogeneous and isotropic three-spaces is
the Robertson--Walker metric,
\be
  \dd s^2 = \dd t^2 - R^2(t) \left[ \frac{\dd \tilde{r}^2}{1-k\tilde{r}^2}
 + \tilde{r}^2 \dd \theta^2 + \tilde{r}^2 \sin^2\theta \,
      \dd \varphi^2 \right] \; , \la{RW}
\ee
where $R(t)$ is the cosmic scale factor, $\tilde{r}$ is a dimensionless scaled
coordinate and the constant $k$ takes values of $+1$, $0$ or $-1$
for three-spaces of positive, zero or negative curvature, respectively.

The full stress-energy tensor, the variation of the action of
${\cal L}_{A,\phi,\psi}$ with respect to the metric, is a complicated object.
However, to obtain the general evolution of the Universe it is enough
to assume that it is given by the simple perfect fluid form
$T^{\mu}_{\;\:\, \nu} \! = \! {\rm diag} ( \varepsilon, -p, -p, -p )$, where
$\varepsilon$ is the energy density and $p$ the pressure.
The independent Einstein equations can in this case be chosen as
\begin{eqnarray}
   \frac{\dot{R}^2}{R^2} & = & \frac{8 \pi G}{3}
      \varepsilon - \frac{k}{R^2}  \; , \la{Friedmann}  \\
  \dd ( \varepsilon R^3 ) & = & - p \, \dd ( R^3 )  \; .  \la{Econs}
\end{eqnarray}
The upper one, or Friedmann equation, gives the time dependence of the scale
factor. The lower equation can also be derived from the local conservation
of energy-momentum of the perfect fluid.

In the case of the very early Universe, when matter behaves like radiation,
pressure and energy density are given by
\be
  p = \frac{1}{3} \varepsilon = \frac{\pi^2}{90} g_* T^4  \la{pressure}  \; ,
\ee
where $g_*$ is the effective number of relativistic degrees of freedom.
Every relativistic bosonic and fermionic degree of freedom increases the value
of $g_*$ with unity and $7/8$, respectively.
The relation (\ref{pressure}) is an idealization, valid for free particles.
Inclusion of the interactions between particles modifies it
slightly even at temperatures far away from any phase transition
[see {\em e.g.}\ \cite{EnqvistSirkka93}].

When the early Universe was at local thermal equilibrium, entropy was
conserved. This can be seen from \eq(\ref{Econs}) using standard
thermodynamics combined with the extreme smallness of all chemical potentials.
The conservation of entropy implies that
\mbox{ $RT =$ constant}, as long as $g_*$ does not change.
The curvature term (the one with $k$) is negligible in the
Friedmann equation~(\ref{Friedmann})  and we can neglect it.
Now it follows from substituting the expression for the energy density
in \eq(\ref{pressure}) to the Friedmann equation
and expressing the gravitational constant as $G = 1 / M_{\rm Pl}^2$ that
\be
  T^2 t = \sqrt{\frac{45}{16 \pi^3}} \,
           \frac{M_{\rm Pl}}{\sqrt{g_{*}}} \; .
    \la{T2t}
\ee
This important equation gives the relation between temperature and time in
a radiation-dominated Universe.

\clearpage

\vsas
\section{First-Order Phase Transitions}   \label{3}

\subsection{Nucleation Rate in Field Theory}   \label{3.1}

Quantum field theory forms the theoretical framework for describing
microscopic relativistic processes at zero temperature.
The theory is elegantly formulated in terms of the
Feynman path integrals. This formulation can be generalized to
non-zero temperatures [\cite{Bernard74}; \cite{GrossPisarskiYaffe81}].
%We will mainly consider a single scalar field.

The thermodynamical properties of a physical system at nonzero temperature
can be calculated from the partition function
\be
  Z = {\rm tr} \; e^{- \beta H} \; ,                         \la{Ztr}
\ee
where $\beta$ is the inverse temperature and $H$ the
Hamilton operator.

The operator $\exp(- \beta H)$ in~\eq(\ref{Ztr}) is analogous to the
quantum-mechanical time evolution operator $\exp(- i H t)$, if one makes the
identification $i t = 1/T$. Indeed, in Euclidean space-time the path-integral
representation for the partition function of a scalar field is
\be
  Z = \int_{\beta-{\rm periodic}} [\dd \phi (\tau,{\bf x})]  \;\;
       \exp \left\{ -\int_0^{\beta} \dd \tau \int \dd^3 x \;
         {\cal L}_E (\phi,\partial \phi) \right\}  \; ,   \la{Zpi}
\ee
where the imaginary time is denoted by $\tau$ and ${\cal L}_E$ is the
Euclidean Lagrangian density. We can see that along the
imaginary-time direction the field $\phi$ propagates over a distance $\beta$.
Because of the trace in the definition of the partition function
in~\eq(\ref{Ztr}), the physical states
%corresponding to the field configurations
have to be identical at ``times'' $0$ and $\beta$.
Thus the scalar field must obey
\be
  \phi (\tau + \beta, {\bf x}) = \phi (\tau, {\bf x})
   \;\;\;\;\;  \forall \; \tau, {\bf x}  \; .
\ee
The subscript `$\beta-$periodic' in~\eq(\ref{Zpi}) refers to this periodic
boundary condition.

For fermion fields the boundary conditions are not periodic but antiperiodic.
This is due to the anticommuting nature of fermions, which is reflected in the
definition of the time-ordering operator. For gauge fields
%the derivation of the partition function path integral is more complicated,
%since
Gauss's law (known in electrodynamics as Maxwell's first law)
must be imposed on physical states. Furthermore,
periodicity of the temporal gauge field $A_0$ is just the most natural choice,
not an automatic consequence.

At this point, two comments related to the partition function are in order.
Firstly, by formulating the theory in Euclidean space-time we loose
information on the real-time evolution of the system. The imaginary-time
formalism is applicable only for describing processes which occur
at or near thermal equilibrium. Most of the time the evolution of
the Universe probably happened very near thermal equilibrium, but not during
the first-order phase transitions.
However, the moment of the onset of a first-order
phase transition can still be calculated using these methods.
Secondly, we have assumed that the chemical potentials for different particle
species vanish.
This is usually justified
when one calculates quantities like pressure or energy
density because of the very small baryon and (presumably) lepton asymmetry
[see {\em e.g.}\ \cite{KajantieKurki-Suonio86}].
%The very small baryon (and presumably lepton) asymmetry justifies
%this assumption, possibly except for extreme baryon number enrichment
%during quark--hadron phase transition (*ch. from HK-S*).

Next, we will discuss the nucleation rate in phase transitions.
We consider a single scalar field, for which the Euclidean
effective action is given by
\be
  S  = \int_0^{\beta} \dd \tau \int \dd^3 x \;
     \left[ \frac{1}{2} \left( \frac{\partial \phi}{\partial \tau} \right)^2
             + \frac{1}{2}( \nabla \phi )^2  +  V(\phi,T)    \right] \; .
        \la{S}
\ee
The potential $V(\phi,T)$ should be understood as an effective potential.
Besides the classical part, it includes both zero-temperature quantum
corrections and finite-temperature thermal corrections. We assume
that the corrections do not change the derivative part of the action.
For a discussion on this point see for instance [\cite{Brahm92}].
%(\cite{KirzhnitsLinde76}; \cite{Brahm92}).

In \fig\ref{fig:potentials}, the qualitative behavior of the potential
is shown. At high temperatures, the only
vacuum of the scalar field model is at $\phi \! = \! 0$.
At some lower temperature, another local minimum starts to develop.
With decreasing temperature, the local minimum soon becomes the true vacuum,
and the vacuum at origin becomes metastable.
At still lower temperatures
%the curvature of the potential at origin changes sign,
the metastable vacuum vanishes,
and the system has again only one vacuum. It is the wall between
the old metastable vacuum and the new true vacuum
that causes the phase transition to be first order.
The thermodynamical transition temperature $T_c$ is the temperature at which
the two vacua are degenerate. In cosmology the system supercools because of
the expansion of the Universe and because the transition rate is only finite,
and therefore
the transition takes place at a temperature somewhat lower than $T_c$.

\begin{figure}[t]
%\pspict{7.5cm}{potentials_noaxnum.ps}
\vspace*{7.5cm}
%klaava:~/matlab/pt/plotpotentials.m
\vspace*{-3.8cm}
\hspace*{2.9cm}
%\special{dvitops: begin fxy7}
%\special{dvitops: origin fxy7}
%\mbox{ \normalsize{$ V(\phi,T) $} }
%\special{dvitops: end}
%\bigskip \bigskip \bigskip \bigskip \bigskip \bigskip
\bigskip \bigskip \bigskip \bigskip \bigskip
%\begin{center}
%\mbox{ ~ ~ \normalsize{$ \phi $} }
%\end{center}
%\vspace*{-0.3cm}
\vspace*{-0.8cm}
%\special{dvitops: rotate fxy7 -90}
\caption[x]{ Qualitative behavior of
  the effective potential for the scalar field, $V(\phi,T)$.
  The potential is plotted at four different temperatures which obey
  $ T_4 < T_3 < T_2 < T_1 $
  and is normalized to vanish at $\phi \! = \! 0$
  for all values of $T$.\la{fig:potentials} }
\end{figure}

We assume that the mechanism for the phase transition is homogeneous
nucleation.
However, one should note that the presence of relic fluctuations or exotic
objects like magnetic monopoles, cosmic strings, black holes or very massive
particles might modify the mechanism.
In the phase transition, bubbles of the stable phase are created. Nucleated
bubbles which are bigger than so-called critical bubbles begin to grow.
When the bubbles grow, the Universe is gradually converted to the new phase.

The decay rate of a metastable vacuum has in the context of statistical field
theory been calculated by Langer [1969]. For relativistic quantum field
theory at zero temperature the decay rate has been evaluated
%At zero temperature, the decay rate of a metastable vacuum in relativistic
%quantum field theory has been calculated in the semiclassical approximation
by Coleman [1977] and Callan and Coleman [1977], and
the same semiclassical methods can be applied also at finite
temperatures
[\cite{Affleck81}; \cite{Linde77}, 1981, 1983; \cite{ArnoldMcLerran87}].
At high temperatures, the mechanism for bubble creation is, instead of
quantum tunnelling, thermal
over-barrier nucleation.
The four-dimensional action can be approximated
with the three-dimensional one:
\be
  S  \approx  \frac{1}{T}  \int \dd^3 x \;
        \left[ \frac{1}{2}( \nabla \phi )^2  +  V(\phi,T)    \right]
   \equiv \frac{S_3}{T}   \; .     \la{S3}
\ee
In the case of weakly first-order phase transitions
%the use of
the three-dimensional
action is at least a very good approximation, if not even exact.

The so-called critical bubble
is a non-vanishing solution of the field equation:
\be
   \frac{\delta S_3}{\delta \phi} \left|_{\phibar} \right.
  = - \nabla^2 \phibar + V'(\phibar,T) = 0    \;  ,  \la{EOM}
\ee
where the prime means derivative with respect to $\phi$.
The boundary conditions are that the derivative of the solution must vanish
at the center of the bubble, chosen to be at the origin,
and that at infinity the solution must be in the metastable vacuum.
It has been shown that the critical bubble is spherically symmetric
[\cite{ColemanGlaserMartin78}].
%More physically speaking, the extremum action $ \bar{S_3} (T) $ is the free
%energy needed to form a critical bubble.
An example of the critical bubble solution $\phibar (r)$ is given later on in
\fig\ref{fig:phiofx} in Subsection~\ref{4.1}.

The decay rate of the metastable vacuum can be calculated from the imaginary
part of the partition function.
The extremum action corresponding to the critical
bubble will be denoted by $ \bar{S_3} (T) $,
and the potential is normalized
as in \fig\ref{fig:potentials} so that the vacuum action of the
high-temperature phase vanishes.
The probability of nucleation per unit time per unit volume is
\be
p(T) = \frac{\omega_-}{2\pi} \left( \frac{ \bar{S_3}(T) }{2\pi T}\right)^{3/2}
  \left| \frac{ \det'[ -\partial^2 + V''(\phibar (r),T) ] }
               { \det [ -\partial^2 + V''(0,T) ] }    \right|^{-1/2}
  \exp \left[ - \bar{S_3}(T) / T \right]   \; ,  \la{Plong}
\ee
where $\det'$ means that the zero eigenvalues resulting from the three
translations of the bubble center should be omitted when calculating the
functional determinant.
The real quantity $\omega_-$ is the (angular) frequency of the unstable mode
of small fluctuations around the bubble solution.

The dominant factors in the nucleation rate~(\ref{Plong})
are the exponential part and the dimensional part of the prefactor. By
prefactor we mean the product of
all the factors that multiply the exponential part.
If one transforms the functional
determinant to dimensionless units, the dimensional quantity that
factorizes out is $M^3 (T)$, where $M(T)$ denotes the mass of
the quadratic term of the potential in the symmetric phase.
%equal to the square-root of the curvature of the potential in the origin.
It has been shown [\cite{BrihayeKunz93}] that
the frequency of the unstable mode can be
%is according to Brihaye and Kunz [1993]
approximated well---at least for some set of parameters of the quartic
%(to be presented in Subsection~\ref{4.1})
potential---with the thin-wall formula
$ \omega_- \! = \! \sqrt{2} / R_{\rm Tcr} $, where $R_{\rm Tcr}$
is the radius of maximal tension of the critical bubble (see Paper~III).
The remaining dimensionless determinant is estimated to differ from unity
at most by a few orders of magnitude. This expectation gains some confidence
from a somewhat similar case, namely from investigations of the sphaleron
transition rate where the corresponding determinant
has been calculated both numerically
[\cite{CarsonMcLerran90}] and analytically [\cite{Carsonetal90}].
Very recently, the fluctuation determinant has been evaluated in the case
of a critical bubble as well [\cite{BaackeKiselev93}].
The result, though renormalization scheme dependent, is
%in the logarithmic scale
not very far from unity for
%values of temperature which are
%not totally different from the bosonic mass of the quartic potential.
realistic values of~$T/M$.

Changing the value of the prefactor with
even several orders of magnitude would not
significantly affect the cosmological nucleation calculations, as we
shall see later on.
%Therefore, the dimensional estimate $T^4$ for
%the prefactor is usually accurate enough, giving for the decay rate
We may rewrite the nucleation rate in the form
\be
  p(T)  = b T_c^4  \, e^{ -\bar{S_3}(T) / T }  \; ,   \la{Pshort}
\ee
where the value of the slowly changing function $b(T)$
%is equal to unity within a few orders of magnitude.
%differs from unity at most by a few orders of magnitude.
is not essential compared with the exponential part.
Besides the simple estimates $T^4$ and $T_c^4$, prefactors like $M^4 (T)$
[\cite{McLerranetal91}] and the Laplace radius of the critical bubble
$R_{\rm Lcr}$ multiplied by some other dimensional factors
[\cite{CsernaiKapusta92}]
have been proposed. ($R_{\rm Lcr}$ will be defined
in \eq(\ref{RcrFcr}) in Subsection~\ref{3.4}.)

\vsass
%\subsection{Phase Transformation}     \label{3.2}
\subsection{Bubble Nucleation and Growth}     \label{3.2} \label{3.3}

A measure telling how the phase transition proceeds is the fraction of space
converted to the new phase.
In cosmology, the formula for the volume fraction remaining in the old
phase given by Guth and Tye [1980] is usually employed
[see also \cite{GuthWeinberg81}].
%\cite{Avrami39} (*ch. ref*) has in the context of crystal growth derived
%an expression for the volume fraction in the old phase, and in cosmology
%a generalized version of it by \cite{GuthTye80} is usually employed
%(see also \cite{GuthWeinberg81}).
For time scales much shorter than the Hubble
time the expansion of the Universe can be neglected. This approximation should
be valid for the whole electroweak phase transition and for the initial stages
of the quark--hadron phase transition. The fraction of space still in the
metastable phase at time $t$ is given by
\be
  f_{\rm ms} (t) =  e^{-I(t)} ;   \;\;\;\;
  I(t) = \int_{t_c}^t \dd t' \, p(T(t')) \, V(t',t)    \;  ,    \la{fGW}
\ee
where $t_c$ is the time when temperature is equal to $T_c$ and
$V(t',t)$ is the volume that a bubble nucleated at time $t'$ occupies
at time $t$.

A different expression for $f_{\rm ms}$
has been proposed by Csernai and Kapusta [1992b].
In their approach, the metastable fraction is given by the integral equation
\be
  f_{\rm ms} (t) =
  1 - \int_{t_c}^t \dd t' \, p(T(t')) \, f_{\rm ms} (t') \, V(t',t)
     \; .    \la{fCK}
\ee

The two expressions
%for the fraction of space in the metastable phase,
(\ref{fGW}) and (\ref{fCK}) coincide in the very beginning of
the phase transition when $ 1 \! - \! f_{\rm ms} \ll 1 $,
and always if the bubbles do not grow. However, they give different results
when growing bubbles begin to overlap. The underlying assumption of
the Guth--Tye formula~(\ref{fGW}) is that when two bubbles come into
contact they simply grow inside each other without any effect
on those parts of the bubbles which do not touch yet. As a consequence
the phase transition is never fully completed in an infinite volume;
there always remains a non-zero fraction of space in the old phase.
The exponentiation of the  naive fraction $I(t)$ takes care of bubble
overlap and of the fact that the volume fraction where new bubbles
can nucleate decreases with time.
The latter feature is explicitly taken into account
in the Csernai--Kapusta formula~(\ref{fCK}), whereas
the volume occupied by two bubbles which have come into contact
with each other is still given as a sum of their individual volumes.

\begin{figure}[t]
\vspace*{7.1cm}
\hspace*{-0.95cm}
%\special{dvitops: import disk$b:<ignatius.tex.phd>tx.ps \the\textwidth 6cm}
%fltxc:~/gle/tx.gle
\hspace*{0.0cm}
\vspace*{-0.9cm}
\caption[a]{World lines of cosmic fluid near a deflagration preceded
  by a shock front. In this idealized figure space-time is 1+1 dimensional,
  and the initial bubble, located at the origin, is infinitesimally small.
  Dashed lines show how two fluid points move. In the old phase,
  temperature is $T_f$ outside the shock, and $T_q$ in the region
  between the shock and the interface. In the new phase temperature is equal
  to $T_h$. Slopes of the lines are the inverse velocities of the shock front,
  the phase boundary (``wall''), and the fluid in the intermediate
  region.\la{fig:tx}}
\bigskip
\end{figure}

Before going further, the concept of bubble in this context should be
clarified. Nucleated bubbles grew probably as deflagrations both in the
electroweak and in the quark--hadron phase transition.
The effect a deflagration bubble has on the cosmic fluid is illustrated
in \fig\ref{fig:tx}. The coordinate system is chosen to be
initially at rest with respect to the cosmic fluid. First, the fluid is hit by
a supersonic shock front which heats it
and accelerates it to the velocity $v_{\rm fluid}$ in
the same direction as the shock. Later on,
a subsonic deflagration front, which is the actual phase boundary,
propagates into the moving fluid transforming it to the new phase.
The fluid is cooled down and stays again at rest.
% ---
Depending on the situation, the bubble surface should be identified
in the calculation as either the phase boundary or as the shock front.
In the latter case the expressions (\ref{fGW}) and (\ref{fCK}) give,
instead of the fraction of space in the old phase, the fraction not touched
by the shocks.

%In reality,
When bubbles begin to touch several things can happen.
If the deflagration fronts expand very slowly,
as is the case in the quark--hadron phase transition
after the system has reheated back to $T_c$ (Subsection~\ref{4.2}),
they rearrange the surface quickly to a spherical shape
so that the total volume
stays approximately constant. However, when the radii
of the bubbles exceed a characteristic
length $R_{\rm fus} (t)$, the rearrangement process is too slow compared
with the cosmic expansion to take place [\cite{Witten84}].
Finally, when the stable phase fills most of the space, the interface has
arranged itself in such a way that the high-temperature phase is located
in isolated droplets.
For slow deflagration fronts in the electroweak phase transition the
process should be similar, except that the characteristic length
$R_{\rm fus} (t)$ is smaller
than the one which follows directly from Witten's argument
since the time scale of the whole phase
transition is much shorter than the Hubble time
(Subsection~\ref{4.1}). The case of shock collisions
%where we should identify the ``bubble'' surface as
%the shock front and not as the actual phase boundary.
%The concept of shock bubble is
which is
relevant for studying the reheating
is qualitatively different, because
shock fronts do not fuse together.
%but reflections may complicate the situation and make their analysis more
%difficult.
%but the analysis is complicated by reflections.

When comparing the Guth--Tye and Csernai--Kapusta expressions for the volume
fraction $f_{\rm ms} (t)$,
one may come to the conclusion that neither of them is perfect. For phase
boundaries, the Guth--Tye formula overestimates the time needed to complete
the phase transition because it assumes that the bubbles grow inside each other
without any interference; the Csernai--Kapusta formula underestimates
the completion time since after two bubbles have
fused together the single volume grows in reality more slowly
than the total volume of two bubbles had they been separated.
For shock fronts, the Guth--Tye formula should be rather accurate
unless reflection plays a significant role in the front collisions.
%[check]
In actual calculations the Guth--Tye formula is easier to handle.
%For some purposes
%a knowledge of the  initial stages of the phase transition is sufficient,
%and in this regime both expressions are accurate enough, as would be the naive
%volume fraction $f_{\rm ms} \! \approx \! 1 \! - \! I$ as well.
We will adopt the Guth--Tye formula~(\ref{fGW}) for further use.

%\vsass
%\subsection{Bubble Nucleation}     \label{3.3}

Let us now inspect more closely that stage of the phase transition
during which most of the bubbles are nucleated.
We will assume that the nucleation is not active any more inside the bubbles.
This assumption is too strong only if we consider slow
deflagrations preceded by shocks,
%identifying the bubbles with the shocks in the calculation,
and if the shocks are, on one hand,
too weak to reheat the plasma enough to stop the nucleation
and, on the other hand,
not so weak that their
effect on the nucleation rate could be neglected.
In the last case, if one
identified the bubble boundary as the deflagration front
and not as the shock, there would be no complications.

%The prefactor in the nucleation rate in \eq(\ref{Pshort}) is almost
%constant compared with the exponentiated nucleation action, as is
%the bubble wall velocity $v$.
The actual period of nucleation is short. During it
%During the period when the bubbles in practice are nucleated,
the prefactor in the expression~(\ref{Pshort}) for the nucleation rate
%as a function of the temperature
changes only slowly compared with the
exponentiated nucleation action, as does the bubble wall velocity $v$.
We may hence approximate them as being constants.
By writing the nucleation rate in the form
\be
  p(t) = p_0 \, e^{-S(t)} \; , \la{Poft}
\ee
we obtain the
following expression for the metastable volume fraction:
\be
  f_{\rm ms} (t) = \exp \left[ - \frac{4\pi}{3} v^3 p_0 \int_{t_c}^t \dd t'
           e^{-S(t')} (t-t')^3  \right] \; .  \la{Iraw}
\ee
The size of the bubble when it is first nucleated is very small,
as will be shown
later on, and therefore it is left out of this expression.
We define $t_f$ as the time when the bubbles occupy a significant
fraction of the space. For simplicity $t_f$ is always called the phase
transition time, even if the bubbles are defined in terms of the shock spheres.
%can as well mean shocks.
%The integral above can be evaluated by noticing
%that it is so rapidly varying that only the values near $t_f$ matter in
%$S(t')$ (\cite{FullerMathewsAlcock88}). We expand the nucleation action as a
%Taylor series,
%\be
%  S(t') = S(t_f) + \beta (t_f - t') + \cdots  \; ; \;\;\;\;
%  \beta  =  - \frac{\dd S }{ \dd t} \left|_{t_f} \right.  \; . \la{Staylor}
%\ee
The above expression for $f_{\rm ms}$ can be evaluated by noticing that
all the essential change in it takes place when
%in relative terms time is very close to $t_f$ and far away from~$t_c$
$(t \! - \! t_c)/(t_f \! - \! t_c)$ is close to unity
[\cite{FullerMathewsAlcock88}].
After expanding the nucleation action as a Taylor series,
the naive fraction $I(t)$ which after
exponentiation gives the fraction of space still in the old vacuum,
$\exp[-I(t)] \! = \! f_{\rm ms} (t)$, can be expressed as
\be
  I(t) = I(t_f) \, e^{-\beta(t_f - t)}   \; ,  \la{Ishort}
\ee
where
\be
  \beta  =  - \frac{\dd S }{ \dd t} \left|_{t_f} \right.  \; . \la{Staylor}
\ee
The positive factor $\beta$ turns out to be an important scale-setting
parameter for the phase transition.

The cosmological phase transition discussed in Paper~II differs qualitatively
from the `normal' cases. It is exceptional in the sense that in thermal
equilibrium the same phase is the true vacuum at all temperatures.
The approximation presented in last paragraph is valid
both for this `exceptional' phase transition and for
the electroweak and the quark--hadron transition when
\be
  \beta \, (t - t_c)  \gg 1  \; . \la{alphacond}
\ee
In normal transitions, the nucleation action is proportional to the square of
$1/(t \! - \! t_c)$, and decreases very rapidly after the critical temperature
$T_c$  has been reached. As a consequence, the validity condition is very
clearly fulfilled in both electroweak and quark--hadron transition:
the left-hand side of
\eq(\ref{alphacond}) is during actual nucleation
bigger than unity by two orders of magnitude.
In the transition considered in Paper~II, the nucleation action behaves
as $\log^{3/2} (1/t)$. (In this case $t_c$ in
\eq(\ref{alphacond}) means the time when the metastable vacua were
created.) Due to the slowness of the decrease of the action,
%the expression in the criterion for validity
the left-hand side of \eq(\ref{alphacond})
is not more than about~6.
However, this value can still be considered to lie within the allowed range.

As in Paper~I, we will
define $t_f$ to be the time when the volume fraction occupied by
the bubbles equals $1/e$, in other words, we put
$I(t_f) \! = \! 1$ in \eq(\ref{Ishort}).
The phase transition time $t_f$ can then be solved from the equation
%for the factor $I(t_f)$,
%in \eq(\ref{Ishort}),
\be
  1 = I(t_f) = \frac{8 \pi v^3}{\beta^4} p(t_f)   \; .   \la{If}
\ee
In principle, the Guth--Tye formula may not be
any more fully valid at $t_f$ due to bubble collisions.
However, the inaccuracy that this causes is negligible when determining
for example how much the system supercools.
%However, the effect of
%redefining $I(t_f)$ as equal to for example
%$0.5$ or $2$ would be negligible for the
%determination of any of the relevant time scales.

As an interlude, let us approximate \eq(\ref{If})
%, which determines the phase transition temperature,
dimensionally as $ p(t_f) t_c^4 \! \approx \! 1 $.
The purpose of this zeroth-order approximation is to give an
estimate of the scales involved. Utilizing the relation between time and
temperature in radiation-dominated Universe, \eq(\ref{T2t}), we obtain
\be
  S(t_f) = \frac{ \bar{S_3} (T_f) }{T_f}
  \approx  4 \log \frac{ M_{\rm Pl} }{T_c}  \; ,  \la{Simple2}
\ee
where $T_f$ stands for the temperature in those parts of space which have not
been affected by the bubbles yet (see \fig\ref{fig:tx}).
Now we can see why uncertainties of several orders of magnitude,
{\em e.g.}\
in the prefactor of the nucleation rate, are not important in cosmology:
the value of the critical nucleation action is very large---as
a result of the very slow expansion rate of the Universe or the weakness of
the gravitational interaction, which is seen as the large value of the
Planck mass.
%At the scales we are interested in, those of
For the transition temperatures of
the electroweak and the quark--hadron phase transition,
the right-hand side of \eq(\ref{Simple2})
is equal to 150--160 or 180--185, respectively.
This also means that the validity of the semiclassical or WKB approximation,
employed in the calculation of the nucleation rate,
should in this regard be on firm footing.

\begin{figure}[t]
\vspace*{5.7cm}
\hspace*{-0.1cm}
%\special{dvitops: import disk$b:<ignatius.tex.phd>tim55.ps 5.7cm 5.7cm}
\hspace*{4.5cm}
%\special{dvitops: import disk$b:<ignatius.tex.phd>tim35.ps 5.7cm 5.7cm}
\hspace*{4.5cm}
%\special{dvitops: import disk$b:<ignatius.tex.phd>tim15.ps 5.7cm 5.7cm}
\hspace*{0.0cm}
\vspace*{-0.6cm}
\caption[a]{A two-dimensional simulation of bubble growth,
  taken from Paper~I\@. The three frames of size $(40 \, v / \beta)^2$
  show the bubble configuration at times $t_f - 4.5/ \beta$,
  $t_f - 2.5/ \beta$ and $t_f - 0.5/ \beta$.
  Bubble collisions were neglected in this simulation,
  but new bubbles were not allowed to nucleate inside
  existing ones.\la{fig:bugrowth}}
\bigskip \medskip
\end{figure}

Let us return to the general, more accurate analysis. As in Paper~I, we
define an effective nucleation rate $\psi (t)$ as follows:
\be
  \psi (t) = p(t) f_{\rm ms} (t) = - \frac{p(t_f)}{\beta}
  \frac{ \dd f_{\rm ms} (t) }{\dd t  }  \; .   \la{psi}
\ee
The decay rate per unit volume per unit time is corrected with
the metastable fraction, since as the phase transition proceeds there is
less space available for the bubbles of the new phase to nucleate.
Integrating the effective nucleation rate we obtain the number of bubbles
in unit volume as a function of time:
\be
  n(t) = \int_{t_c}^t \dd t' \psi (t')
  = \frac{p(t_f)}{\beta} \left[ 1 - f_{\rm ms} (t) \right]  \; .  \la{n}
\ee
The number density increases linearly with the fraction of space in the
low-temperature vacuum.

A good estimate for the final number density of
bubbles $n_{\rm final}$ is obtained by setting $f_{\rm ms} \! = \! 0$
in \eq(\ref{n})
and using \eq(\ref{If}).
%The true number density is expected to be somewhat
%smaller since the Guth--Tye formula~(\ref{fGW}), which we have used,
%slightly overestimates the duration of the phase transition.
%However, this error is expected not to be significant.
%On the other hand,
If several collisions of
the shocks are needed to produce notable reheating
%(the case discussed a few paragraphs ago),
(Subsection~\ref{4.2}),
the true final number density is
%correspondingly
somewhat larger
than the one obtained from \eq(\ref{n}).
%The inaccuracies reflect also on
The average
distance of nucleation centers $R_{\rm nucl}$, defined as
$n_{\rm final} \! = \! 1/R_{\rm nucl}^3$, is given by
\be
  R_{\rm nucl} = 2 \pi^{1/3} \frac{ v }{ \beta }  \; ,     \la{Rnucl}
\ee
where $\beta$ is defined in \eq(\ref{Staylor}).
%If the velocity of the bubble wall is small, the average nucleation
%distance is small too, since there is more time for new bubbles to nucleate.
Increase in the rate of change of the action means a decrease in the
nucleation distance because then more bubbles nucleate during a given time
interval after the nucleation has effectively been turned on.

In both the electroweak and the quark--hadron phase transition there were
a vast number of bubbles nucleated inside the horizon.
This was partially due to
the rapid change in the nucleation action, and partially due to the smallness
of the dimensionless ratio $\sigma ^{3/2} / L \sqrt{T_c}$, where $\sigma$ is
the interface tension and $L$ the latent heat of the transition
%of the parameter $C$, to be defined in
(see \eq(\ref{Stw}) in Subsection~\ref{3.4}).
On the other hand, in the case of the phase transition suggested in
Paper~II the nucleation action decreases so slowly that there were
only a few hundred nucleated bubbles inside one horizon.
%The condition that there should be at least one bubble inside the horizon
%is very clearly fulfilled both in the electroweak and in
%the quark--hadron phase transition (see Section~\ref{4}).
% --- No such condition exists, I think!

Let us now consider the distribution of bubble sizes.
%(\cite{TurnerWeinbergWidrow92}).
This has been recently discussed by
Turner, Weinberg and Widrow~[1992].
\begin{figure}[t]
%\pspict{7.5cm}{busizedistr.ps}
\vspace*{7.5cm}
%klaava:~/matlab/pt/plotbusizedistr.m
%fltxa:~/ps> matps -X -y -a 0 -b 1 -f 1.6 -l 4 busizedistr.ps
\vspace*{-3.8cm}
%\hspace*{2.9cm}
\hspace*{2.5cm}
%\special{dvitops: begin fxy7}
%\special{dvitops: origin fxy7}
%\mbox{ \normalsize{$ g_t (\rho) \: / [p(t_f)/v] $} }
%\special{dvitops: end}
%\bigskip \bigskip \bigskip \bigskip \bigskip \bigskip
\bigskip \bigskip \bigskip \bigskip \bigskip
%\begin{center}
%\mbox{ ~ ~ \normalsize{$ \beta \, ( t_f - t + \rho /v ) $} }
%\end{center}
%\vspace*{-0.3cm}
\vspace*{-0.8cm}
%\special{dvitops: rotate fxy7 -90}
\caption[x]{ Bubble size distribution.
  The area to the right of the right vertical line is the bubble size
  distribution at time $t \! = \! t_f \! - \! 3/ \beta$,
  and the area to the right of the left vertical line is the bubble size
  distribution at time $t_f$.
  In other words, the vertical lines correspond bubbles of zero-radius
  at those times.\la{fig:busizedistr} }
\bigskip
\end{figure}
The bubble size distribution is easily obtained
after realizing that a bubble which has a radius $\rho$ at time $t$,
was nucleated at time $t \! - \! \rho / v$.
Here we made again the approximation that at nucleation the bubble radius is
so small that it can be neglected (see Subsection~\ref{3.4}).
The distribution of bubble sizes is hence given by
\be
  g_t (\rho)
%= \frac{\dd n}{\dd \rho} \left|_t \right.
  = \frac{1}{v} \, \psi ( t - \rho / v )   \; .    \la{gtrho}
\ee
The number of bubbles in unit volume at time $t$ with radius between
$\rho$ and $\rho \! + \! \dd \rho$ is equal to $g_t(\rho) \dd \rho$.
More explicitly, the bubble size distribution is given  by
\be
  g_t (\rho) = \frac{p(t_f)}{v} \: e^{-u} \, \exp \left[ - e^{-u} \right]
     \; ; \;\;\;\;  u = \beta \, ( t_f - t + \rho / v )  \; .  \la{gtrho2}
\ee
When inspecting the spatial scale of density inhomogeneities,
the distribution of distances between
nucleation sites of bubbles near each other
could likewise be used.  Meyer \etal [1991] have studied
%that distribution
it extending the naive fraction approximation
for the whole nucleation period.

The distribution of bubble sizes given in \eq(\ref{gtrho2}) is presented
in \fig\ref{fig:busizedistr}.
%When interpreting the figure,
%one must note that bubble with zero radius is the smallest one.
When time $t$ is in the vicinity of the phase transition time~$t_f$,
our approach loses its
reliability. Therefore the distribution at time $t_f$,
the area to the right of the left vertical line in the figure,
may well be somewhat distorted near the origin,
whereas the other distribution
corresponding to $t = t_f - ( 3/ \beta )$
can be expected to be quite accurate.
The bubble size distribution (\ref{gtrho2})
is universal in the sense that if the validity
of the assumptions is equally good in two different phase transitions, the
appropriately scaled distributions should look the same.

\vsass
\subsection{Thin-Wall Limit}     \label{3.4}

In the so-called thin-wall limit, or small relative
supercooling limit, the equations for the
amount of supercooling and the distance between nucleation sites
can be expressed in a simple form and solved with a good accuracy.
Let us consider a thin-walled bubble of radius $R$.
Let the whole interior of the bubble be in
the vacuum of the new phase and the whole exterior in the old vacuum,
and let the interface between be infinitely thin.
%with an infinitely thin interface in between.
The free energy density
difference between the two phases is,
in the absence of any relevant conserved charge,
equal to minus the pressure difference~$\Delta p$.
By difference we mean the value of the quantity in the low-temperature phase
minus that in the high-temperature phase.
Hence the free energy of the bubble is
\be
  F(R) = - \frac{4 \pi}{3} \Delta p \, R^3 + 4 \pi \sigma R^2  \; , \la{Fexp}
\ee
where $\sigma$ is the interface
%(surface)
tension.
A generalization of this expansion is discussed in detail in Paper~III\@.
The radius of the critical bubble and the corresponding free energy density
are found by maximizing $F(R)$ with respect to $R$, with the result
\be
  R_{\rm Lcr} = \frac{2 \sigma}{\Delta p} \; , \;\;\;\;
  F_{\rm cr} = \frac{16 \pi}{3} \frac{\sigma^3}{(\Delta p)^2} \; . \la{RcrFcr}
\ee
%The subscript L denotes that this is Laplace's definition for the bubble
%radius.
Above $R_{\rm Lcr}$ is Laplace's definition for the radius of the critical
bubble.
The probability of fluctuation in which a critical bubble is formed
is proportional to the Boltzmann factor $\exp ( - F_{\rm cr}/T )$.

For the normal phase transitions the pressure difference can be written as
$ \Delta p(T) \! \approx \! L (1 \! - \! T/T_c)$, as given
in Section~\ref{4} in \eq(\ref{Deltap}).
%Comparing with the Boltzmann factor,
%this leads to the following thin-wall nucleation action,
Using \eq(\ref{RcrFcr}), the nucleation action appearing in \eq(\ref{Poft})
can be expressed as follows:
\be
  S(T) = \frac{C^2}{(1-T/T_c)^2} \; ; \;\;\;\;
  C = 4 \sqrt{\frac{\pi}{3}} \frac{\sigma^{3/2}}{L \sqrt{T_c}}  \; , \la{Stw}
\ee
where $L$ is the latent heat of the transition.
%This result is well known in the classical nucleation theory.
In the case of the exceptional phase transition discussed in Paper~II,
one had to use the exact expression for the pressure difference.
Thanks to the simple form of the expression
this would not give rise to any difficulties in the thin-wall calculation.
In the rest of this Subsection, we will analyze only the normal phase
transitions.

Utilizing \eq(\ref{T2t}) and using the fact that $T$ is very close to $T_c$,
we can write the derivative $\beta$ of the action in \eq(\ref{Staylor})
%multiplied by $t_c$
as $ t_c \beta \! = \! S(T_f) / (1 \! - \! T_f/T_c) $.
Substituting the expression~(\ref{Pshort}) for the nucleation rate,
the rightmost part of \eq(\ref{If}) then becomes
\be
  \frac{ 8 \pi v^3 (t_c T_c)^4 b }{ (t_c \beta)^4 } \,
  e^{ - C^2/(1-T_f/T_c)^2 }
 = \exp \left[ A - \frac{C^2}{(1-T_f/T_c)^2}
     - 8 \log \frac{C}{(1-T_f/T_c)^{3/2}}  \right]    \; ,  \la{Iftw}
\ee
where
\be
  A = 4 \log \left( \frac{M_{\rm Pl}}{T_c} \right)
     + 2 \log \left( \frac{45}{16 \pi^3 g_* \! \left|_{t_c} \right. } \right)
     + \log ( 8 \pi v^3 b )      \;  , \la{A}
\ee
and $b$ is the dimensionless factor
%in the expression for the nucleation rate
introduced in \eq(\ref{Pshort}).
The phase transition temperature is obtained by demanding
that the argument of the exponential function on the right-hand side
of \eq(\ref{Iftw}) vanishes.
%It would be to inaccurate to neglect the logarithmic term; however,
A good accuracy is achieved by iterating the resulting equation once.
%The amount of supercooling that results is
This procedure gives the following expression for the amount of supercooling
[\cite{Kajantie92}]:
\be
  1 - \frac{T_f}{T_c} = \frac{C}{ \sqrt{\tilde{A}} }  \; ; \;\;\;\;
  \tilde{A} = A - 4 \log ( A^{3/2} /C )    \; .   \la{Tftw}
\ee
The phase transition temperature $T_f$ given by this equation
should be close to the equilibrium transition temperature $T_c$,
otherwise the approximation used is invalid.

In the expression for $A$ in \eq(\ref{A}) the first term is by far
the dominant one, while for $\tilde{A}$ the second term does have some
significance; for example in realistic estimates for the quark--hadron
transition its value is 30\% of that of the first term in the
expression~(\ref{Tftw}).
The physical reason for such a large value of this correction term
is that the growth
time of bubbles, although tremendously longer than
the time scale of microscopic interactions,
is still very far from the Hubble time.
%still is significantly shorter than the Hubble time.

Now we can determine the average distance between nucleation centers. We
compare it with the horizon radius $R_{\rm hor} \! = \! 2 t_c$:
\be
  \frac{R_{\rm nucl}}{R_{\rm hor}} = \frac{ \pi^{1/3} v C }{ \tilde{A}^{3/2} }
  = 6.00 \, \frac{v}{ \tilde{A}^{3/2} } \frac{ \sigma^{3/2} }{ L \sqrt{T_c} }
     \; .   \la{Rnucltw}
\ee
The later a phase transition takes place, the more bubbles there tends to be
inside the horizon. For example, in a phase transition at the grand unified
scale there should be less bubbles created than in a phase transition
at the QCD scale.
Of course, in actual cases the dependence
of $R_{\rm nucl} / R_{\rm hor}$ on the microscopic ratio $\sigma^{3/2} / L$
%in \eq(\ref{Rnucltw})
may well cancel the dependence on the factor $\log ( M_{\rm Pl} / T )$
related to the cosmic background.

It is also interesting to calculate the radius of the critical bubble
$R_{\rm cr}$, {\em i.e.},
the bubble size immediately after the nucleation. It can be determined from
Laplace's relation~(\ref{RcrFcr}) and is given by
\be
  R_{\rm cr} (T_f) \, T_c
%= \frac{ 2 \sigma }{ \Delta p(T_f) } T_c
  = \sqrt{ \frac{3}{4 \pi} \frac{ \tilde{A} }{ \sigma / T_c^3 } }
       \; .   \la{Rcrtw}
\ee
%where $\Delta p$ is the pressure difference between the two phases.
Here $R_{\rm cr}$ is compared with $1/T_c$, which gives
the length scale for microscopic processes.
%This thin-wall calculation gives an upper limit for
%the real radius of the critical bubble.
% --- No, its a lower limit!

{}From \eqs(\ref{Rnucltw}) and~(\ref{Rcrtw})
we can observe that a natural length scale for $R_{\rm nucl}$ is
roughly set by the horizon
radius, and for $R_{\rm cr}$ by the microscopic length $1/T_c$.
We can thus conclude that the
bubbles are indeed microscopic when first nucleated, but grow to
a macroscopic size before colliding with other bubbles
(unless the interface tension is extremely small).

\clearpage

\vsas
\section{Phenomenology of Cosmological Phase Transitions}  \label{4}

In this Section the general methods
%presented earlier in the Thesis
described above
are applied to the case of the electroweak and
the quark--hadron phase transition.
%The aim is to achieve, at least partially, a quantitative understanding
We present a partially quantitative description
of the different events that took place during these cosmological transitions.

%It should be noted here that
We consider in this introductory review part
only the first-order electroweak and quark--hadron phase transitions
and their cosmological consequences,
but whether these transitions in Nature really are first order
is by no means known with certainty.
A second-order phase transition would have had less
cosmological significance.
%(An exception might be the scenario presented
%by \cite{BrandenbergerDavis93} in which the net baryon number was generated by
%electroweak strings, supposing they were stable.)

A fundamental ingredient for the study of a first-order phase transition
is the thermodynamical equation of state. The absence of chemical
potential simplifies the thermodynamics, since the physical
quantities become functions of only one variable, the temperature.
The pressure $p(T)$, now equal to minus the free energy density,
is taken as the basic quantity.
%It is given by \eq(\ref{pressure}) when the effective number of
%relativistic degrees of freedom stays constant.

When the temperature is in the vicinity of the
thermodynamical transition temperature $T_c$, the pressure difference between
the stable and the metastable phases can be expanded as
\be
  \Delta p (T) = L \left( 1 - \frac{T}{T_c} \right) + \cdots \; . \la{Deltap}
\ee
The quantity $\Delta p (T)$ is defined as the pressure in the low-temperature
phase minus the pressure in the high-temperature phase.
However, this expansion is not valid for the exceptional phase transition
suggested in Paper~II because no critical temperature $T_c$ exists there.

Given the pressure $p(T)$, the entropy and the energy density can be
determined from
\be
  s(T) = \frac{\dd p}{\dd T}  \; , \;\;\;\;
  \varepsilon (T) = Ts - p  \; . \la{epsands}
\ee
These relations hold in both phases.

\vsass
\subsection{The Electroweak Phase Transition}   \label{4.1}

%It is believed that the electroweak symmetry was broken in a phase
%transition that took place via Higgs mechanism
%at a temperature of the order of 100~GeV.
Although the concept of symmetry restoration at high temperatures has been
known for long [\cite{Kirzhnits72}; \cite{KirzhnitsLinde72};
\cite{DolanJackiw74}; \cite{Weinberg74}], the electroweak phase
transition has remained poorly understood until recent times.
The interest in it was renewed some time after the
observation that nonperturbative processes, mediated by the
sphalerons,
have a significant effect on the baryon asymmetry of the Universe
[Kuzmin, Rubakov and Shaposhnikov~1985]. Since then, several authors
have studied the dynamics of the first-order electroweak phase transition
[\cite{McLerranetal91}; \cite{Turok92}; Paper~I; Anderson and Hall~1992;
\cite{Dineetal92}; Liu, McLerran and Turok~1992;
Carrington and Kapusta~1993].

The minimal standard model of electroweak interactions
contains one scalar doublet. The
order parameter of the electroweak phase transition is the field
corresponding to
the low-temperature physical Higgs particle,
usually taken as the real part of that
component of the doublet which acquires a vacuum expectation value.

The Higgs boson has never been experimentally detected, and guesses for its
mass cover a wide range.
%It is even possible that it does not exist;
Even its existence may be questioned;
for instance, a condensate of heavy quarks could effectively act as
a Higgs particle [\cite{Lindner92}].
Futhermore, it might be
that the real electroweak theory contains more scalar fields than just one
doublet.
In the case of two doublets a finite temperature potential of the same
form as in the case of one doublet
%(to be presented in \eq(\ref{Veikr}))
can be used as a reasonable approximation
for that combination of scalar fields which
%first becomes massless
drives the transition
[\cite{McLerranetal91}].

The first-order electroweak phase transition is commonly described by
the following effective quartic potential [\cite{Linde83}]:
\be
  V( \phi, T) = \frac{1}{2} \gamma (T^2 - T_0^2) \phi^2
     - \frac{1}{3} \alpha T \phi^3 + \frac{1}{4} \lambda \phi^4 \; , \la{Veikr}
\ee
where the order parameter field $\phi (x)$ is a real scalar function.
The qualitative (but not quantitative)
behavior of this potential is shown
in \fig\ref{fig:potentials}. The thermodynamical transition temperature is
\be
  T_c = \frac{T_0}{\sqrt{ 1-\frac{2}{9}\frac{\alpha^2}{\lambda \gamma} }}
   \; , \la{Tc}
\ee
and the lowest temperature where the symmetric vacuum can exist is $T_0$.

The potential $V(\phi,T)$ could in principle be derived from the full
microscopic theory. The question how to best determine the effective potential
beyond the naive one-loop level is currently under an active study.
One method is to employ effective three-dimensional theories, with
[\cite{Bunketal93}; Kajantie, Rummukainen and Shaposhnikov 1993;
\cite{Farakosetal93}]
or without [\cite{Shaposhnikov93}] lattice Monte Carlo simulations.
These studies
seem to indicate that the electroweak phase transition would not be as weakly
first order as expected on the basis of perturbative analysis.
(There are arguments telling that the existence of
supersymmetry could also make the transition more strongly first order
[\cite{EspinosaQuirosZwirner93}].)

The point of view taken here is that the potential in \eq(\ref{Veikr})
should be regarded as a phenomenological one, valid in the vicinity
of $T_c$. The parameters $T_0,$ $\gamma,$ $\alpha$ and $\lambda$
are to be chosen so that the potential quantitatively correctly describes the
phase transition.
In principle, the values of these parameters could be determined
by experiment or
observation. In practice, first-principles calculations, even if not fully
satisfactory, are employed
%at least in giving a hint
to obtain an estimate
of the relevant scale of the parameters.

Next, we will discuss the equation of state near the electroweak phase
transition. Let us inspect two configurations
where the field $\phi$ is spatially uniform with a value corresponding
either to the minimum at the origin or the other minimum of $V(\phi,T)$.
Since the potential vanishes at origin,
the difference in the free energy densities
between these two configurations is equal to the value
of the effective potential at the other minimum.
%the position of which is denoted by $v(T)$.
We denote it as $V(v(T),T) \! \equiv \! - \! \tilde{B} (T)$,
where $v(T)$ is the value of the field at the other minimum.
This leads to the following expressions for the pressure (Paper~I):\footnote{
 For simplicity the subscripts `q' and `h', adopted from the quark--hadron
 transition, are used also here to denote the high- and low-temperature phase,
 respectively.}
\be
  p_q (T) = a T^4  \; , \;\;\;\;
  p_h (T) = a T^4 + \tilde{B} (T)  \; . \la{pEW}
\ee
The same factor $a$ is used in both radiative terms,
because at $T_c$ the pressures of both phases
must be equal and $\tilde{B} (T)$ vanishes.
The term $\tilde{B} (T)$ was attached to the low-temperature phase
since both $\tilde{B} (T)$ and the low-temperature phase exist
only up to a certain temperature somewhat above~$T_c$.
Moreover, one might add a constant $- \! \tilde{B} (0)$ to each of the phases
in order to make $p_h (T)$ vanish at zero temperature.
However, this is unnecessary
%the constant can as well be dropped out
since the above equation
of state is invalid at low temperatures anyway (Paper~I).

{}From the relation~(\ref{pressure}) one observes that the scale for the
constant $a$ in \eq(\ref{pEW}) is apart from the factor $\pi^2/90$
set by $g_*$, the effective number of relativistic degrees of freedom.
At very high temperatures all the particles of the minimal standard
model contribute to~$g_*$ giving
\be
  g_* =
%\! \left|_{ T \gg T_c^{\rm EW} } \right. =
   \underbrace{2}_{\gamma} + \underbrace{3 \times 2}_{ A_{\mu}^{\rm weak} }
   + \underbrace{8 \times 2}_{g} + \underbrace{4}_{\Phi}
   + \frac{7}{8} \left[ \underbrace{3 \times 4}_{e,\mu,\tau}
      + \underbrace{3 \times 2}_{\nu}
      + \underbrace{6 \times 3 \times 4}_{q} \right]
  = 106.75  \; . \la{gstd}
\ee
At the phase transition all three weak gauge bosons increase their degrees
of freedom with unity by eating the altogether three
would-be Goldstone bosons from the scalar doublet.
%, and at the same time the fourth component of the doublet becomes massive.
This does not change the value of $g_*$ significantly,
because slightly below $T_c$ the masses
of the gauge bosons and the physical Higgs
particle are presumably small compared with the temperature.
However, the top--quark,
the other yet undetected particle predicted by the minimal standard
model, might decrease the value of $g_*$
%increase the change of the value of $g_*$ (the jump in it)
significantly---the contribution from a single quark to the value of $g_*$
%in \eq(\ref{gstd})
is about 10\%.

At the phase transition, the jump in $g_*$, folded
together with the change in the interactions
causing deviation from the
radiative free-gas behavior, produces latent heat of the transition.
For the equation of state~(\ref{pEW}) the latent heat is given by
$L \! = \! - \! T_c \tilde{B} '(T_c)$, and its value is hence reflected in
the parameters of the potential~(\ref{Veikr}).
One should note that latent heat $L$ includes not only the effect of
the Higgs particle but also of
all the other particles, since in our order
parameter model all the other fields have been integrated out.

\begin{figure}[t]
%\pspict{7.5cm}{phiofx.ps}
\vspace*{7.5cm}
%klaava:~/matlab/s3act_cray/plotphiofx.m
%fltxa:~/ps> matps -X -y -a 0 -b 1 -f 1.6 -l 4 phiofx.ps
\vspace*{-3.8cm}
\hspace*{2.5cm}
%\special{dvitops: begin fxy7}
%\special{dvitops: origin fxy7}
%\mbox{ \normalsize{$ \phibar (r') \, / v(T) $} }
%\special{dvitops: end}
\bigskip \bigskip \bigskip \bigskip \bigskip
%\begin{center}
%\mbox{ ~ ~ \normalsize{$ r' $} }
%\end{center}
\vspace*{-0.8cm}
%\special{dvitops: rotate fxy7 -90}
\caption[x]{ Two critical bubbles.
  Dotted line is the extremum solution for $\lbar \! = \! 0.9$
  and solid line for $\lbar \! = \! 0.6$.
  The dimensionless radial variable is $r' \! = \! M r$, where $M$ stands for
  the bosonic mass in the potential,
  $M^2 (T) \! = \! \gamma (T^2 \! - \! T_0^2)$.\la{fig:phiofx} }
\bigskip
\end{figure}

As we have now specified the potential $V(\phi,T)$
we will shortly return to the nucleation process.
It is convenient to express the temperature
dependence of various quantities by using the function $\lbar(T)$,
\be
  \lbar (T) =  \frac{9}{2} \frac{\lambda \gamma}{\alpha^2}
                 \left( 1 - \frac{T_0^2}{T^2} \right) \; ,    \la{lbar}
\ee
which satisfies $\lbar (T_0) \! = \! 0$, $\lbar(T_c) \! = \! 1$.
The nucleation action is written in terms of the function
$\lbar (T)$ as follows:
\be
  S ( \lbar (T) )
%= \frac{ F_{\rm cr} (T) }{ T }
  = \frac{2^{9/2} \pi}{3^5} \frac{\alpha}{\lambda^{3/2}}
    \frac{ f(\lbar) }{ (1-\lbar)^2 } \; .   \la{fdef}
\ee
The function $f(\lbar)$ has been determined in
Papers I and~III by solving numerically the field equation~(\ref{EOM})
for the potential~(\ref{Veikr}). It is a smoothly behaving function with
the special value $f(1) \! = \! 1$. In \fig\ref{fig:phiofx},
the bubble solution of the field equation is shown for two
illustrative values of $\lbar$.
For $\lbar \! = \! 0.9$ the solution somewhat resembles a thin-walled bubble,
but for $\lbar \! = \! 0.6$ the bubble core is far away from the true
vacuum.

{}From \eq(\ref{If}) one can solve for $t_f$,
the age of the Universe when the bubbles had filled the space.
%substituting unity for~$I(t_f)$.
Explicitly, the equation determining
the amount of supercooling is,
in analogy with demanding that the right side of \eq(\ref{Iftw}) equals unity,
as follows:
\be
  A  -  S (\lbar _f)  -  4 \log \left[ \frac{1}{ (T_c/T_0)^2 - 1}
      \frac{\dd S}{\dd \lbar} \left|_{\lbar _f} \right.  \right]
  =  0 \; , \la{Ifacc}
\ee
where $A$ is given by \eq(\ref{A}) and $\lbar _f \! \equiv \! \lbar (T_f)$.
The thin-wall limit, or small relative supercooling scenario,
considered in Subsection~\ref{3.4}
follows as a special case if the solution $\lbar_f$ of
the above equation is close to unity.
For example, the expression~(\ref{Stw}) for the nucleation action follows
from \eq(\ref{fdef}) in the limit $\lbar _f \! \rightarrow \! 1$.

Let us now inspect in more detail how the cosmological electroweak
phase transition is assumed to have proceeded.
The sequence of events is illustrated in
\fig\ref{fig:EW}.
The numerical values of various quantities presented
in the figure correspond to the following values of the parameters:
\be
\begin{array}{rclcrcl}
  T_c  &   = & 100 \: \, {\rm GeV} , & &
  \gamma & = & 0.1309  \: ,  \\
  \alpha & = & 0.0162  \: , & &
  \lambda &= & 0.0131  \: .
\end{array}  \la{EWparams}
\ee
The values of $\gamma$, $\alpha$ and $\lambda$
are those obtained by Huet \etal [1993] by substituting the
zero-temperature masses $M_t \! = \! M_W$,
$M_h \! = \! 40 \: {\rm GeV} \! \simeq \! M_W /2$ into
the perturbatively evaluated
%improved 1--loop
effective potential.
Here the unrealistically low mass for the Higgs--particle
is more crucial than that for the top-quark,
as will be discussed shortly.
%commented in the paragraph below.
The parameter values (\ref{EWparams}), used here
just for illustration,
imply that the phase transition is very weakly first order.
This is indicated by the small latent heat (\mbox{$= 0.086 \, T_c^4$})
and the long correlation length (\mbox{$= 15.0 /T_c$} in both phases at $T_c$).

Trust in the perturbative calculations leads to complications here.
%A Higgs mass of approximately $40 \: {\rm GeV}$ is
The Higgs mass used among others by Huet {\em et al.\/}~[1993] is
close to the upper limit of the allowed region
if one requires that the baryon asymmetry of the Universe was
created in the electroweak transition.
A necessary condition for the electroweak baryon asymmetry generation is that
the sphaleron transitions should have been frozen out in the low-temperature
phase already at $T_c$.
For that to have happened the Higgs scalar must not be heavier than about
40~GeV according to perturbative analysis [{\em e.g.}, \cite{Huetetal92}].
%This requirement is necessary in order to have
On the other hand, the Higgs mass of such a low value
seems to be ruled out by LEP experiments [\cite{Davier92}].
As a solution to this dilemma some authors have assumed the
existence of additional scalar fields.
%since in extended models
In multi-Higgs cases
the parameter $M_h$
%used in determining the effective potential
is not the true zero temperature mass of a physical Higgs particle
and is hence not constrained by the mass measurements
[see \cite{LiuMcLerranTurok92}, and references therein].

In our phenomenological approach the parameter values
quoted in \eq(\ref{EWparams}) do not pose any problems.
Even within the minimal standard model these values
do not imply a Higgs mass which were ruled out experimentally,
as long as one does not try to extend the validity of the potential
down to zero temperatures.

%The parameter values above are used just for illustration,
%and they should by no means be taken as final values.
%As discussed earlier, the transition
%could be more strongly first order, which in turn would reflect in the
%parameter values.

\begin{figure}[t]
\vspace*{9cm}
\hspace*{-0.25cm}
%\special{dvitops: import disk$b:<ignatius.tex.phd>ew.ps \the\textwidth 12cm}
%Sohvi:Die Hard 40/Janne/Kuvia/PhD/EW
\hspace*{0.0cm}
\vspace*{-0.5cm}
\caption[a]{Schematic 1+1 dimensional figure of the electroweak
  phase transition.
  Grey area presents the high-temperature and
  white the low-temperature phase, and the separate pictures at right give
  two-dimensional snapshots of the transition. The numbers are for
  the parameter values in \eq(\ref{EWparams}).
  The idea for this figure is borrowed from
  Rummukainen [1990].\la{fig:EW}}
\bigskip
\end{figure}

In \fig\ref{fig:EW}, three essentially different time scales can be seen.
One is the Hubble time at the transition. Another is the duration of time
the Universe stayed in the supercooled metastable state,
$\Delta t_{\rm sc} = t_f - t_c$.
The phase transition time $t_f$
%This
can be determined from \eq(\ref{Ifacc}) by utilizing
for the nucleation action the numerical function given in Paper~III\@.
At the nucleation, the value of the action turns out to be only 106 instead of
the naive value which according to \eq(\ref{Ifacc}) equals
$A$ (=~145), {\em i.e.}, the
last term in that equation does have some significance.
The shortest time scale appearing in \fig\ref{fig:EW} is the growth
time of the shocks, defined as $\Delta t_{\rm sh} \! = \! R_{\rm nucl} /v$.
The average distance of nucleation centers $R_{\rm nucl}$
is defined in \eq(\ref{Rnucl}).
In numerical calculations the dimensionless prefactor
$b$ of the nucleation rate in \eq(\ref{Pshort}) was taken to be unity
and the free-gas value $v \! = \! 1/ \sqrt{3}$ was used for the shock velocity.

The phase transition was completed at time $t_e$ when the
actual bubbles of new phase, expanding as deflagrations behind the shocks,
had met and coalesced fully. (Detonations, the other type of explosive
processes, would require much more supercooling
than \eq(\ref{Ifacc}) indicates for the present parameters.)
The growth time of the bubbles of the new phase
cannot be solved, because the velocity of
the deflagration front is not known. However, it seems probable that
the difference of the velocities of the deflagration and shock fronts
was clearly less than one order of magnitude
[\cite{IgnatiusKajantieKurki-SuonioLaine93}],
which indicates that the growth time of
deflagrations did not differ drastically from $\Delta t_{\rm sh}$.
This also means
that the inaccuracies in the estimates of the time scales are
not significant ({\em cf.}\ Subsection~\ref{4.2}).

Mutual collisions of the shocks reheat the system to a certain
temperature~$T_{\rm rh}$.
As will be demonstrated below,
the reheating temperature $T_{\rm rh}$ was in the electroweak phase transition
less than the critical temperature $T_c$.
Because of this and the fastness of
the phase transition, $T_{\rm rh}$ can be estimated by going to the
extreme case where the whole space were converted instantaneously
from the old to the new phase at time~$t_f$.
\mbox{DeGrand} and Kajantie [1984] called this scenario ``abrupt transition''.
%(\cite{DeGrandKajantie84}).
The reheating temperature can be obtained from the equation
\be
  \varepsilon_h (T_{\rm rh}) = \varepsilon_q (T_f)  \; . \la{reheat}
\ee
For the parameter values given in \eq(\ref{EWparams})
the reheating turns out to be only 13\% on the scale
where 100\% would mean reheating back to the critical temperature.\footnote{
 In condensed matter physics a first-order phase transition with less
 than 100\% reheating is called ``hypercooled'' instead of supercooled
 [see for example \cite{LeggettYip90}].}
The fact that the critical temperature $T_c$ is not reached during the
reheating is crucial for the
scenario of baryon asymmetry generation discussed by
Liu, McLerran and Turok [1992].
If the reheating temperature $T_{\rm rh}$ had been close to $T_c$,
the velocity of deflagration fronts would have decreased substantially
(see Subsection~\ref{4.2}). In that case
the baryon number produced
in the front would have had time to diffuse to the old phase,
where it would have been washed out by the sphalerons.

At the final stage of the phase transition the shrinking droplets of the old
phase produced rarefaction waves; however,
their effect was shadowed by the presence of the remnants of shocks.
Recently Huet \etal [1993] have demonstrated
that the deflagration fronts did not
develop any instabilities while expanding.
Thus the length scale of inhomogeneities is given by
the usual expression~(\ref{Rnucl}) for the average distance
of nucleation centers $R_{\rm nucl}$. However,
it is hardly probable that the density inhomogeneities produced in the
electroweak phase transition could have been of any importance.
%had much significance.
For instance, for
the parameters (\ref{EWparams}) one can estimate
the maximal relative pressure differences
to be only $\Delta p /p \approx  4 \times 10^{-5}$.
In comparison, for the storms in the atmosphere of Earth
this quantity can be three orders of magnitude larger.

In \fig\ref{fig:Tt} the behavior of temperature versus time
both in the electroweak and in the quark--hadron transition is shown
schematically in a log--log plot. From the figure we can see that the
electroweak phase transition did not last long. Afterwards,
the usual relation between temperature and time,
given in \eq(\ref{T2t}), became soon valid again.

\begin{figure}[t]
%\vspace*{7.1cm}
%\hspace*{-0.95cm}
%\special{dvitops: import disk$b:<ignatius.tex.phd>tx.ps \the\textwidth 6cm}
%\pspict{6.8cm}{disk$b:<ignatius.tex.phd>tt.ps}
\vspace*{6.8cm}
%(was 7.2cm)
%Sohvi:Die Hard 40/Janne/Kuvia/PhD/Tt
%\hspace*{0.0cm}
%\vspace*{-0.9cm}
\caption[a]{ Relation between temperature and time in the electroweak (EW) and
  in the quark--hadron (QH) phase transition.
  Dotted curve denotes the reheating period during which temperature was
  far from being spatially uniform.\la{fig:Tt}}
\medskip
\bigskip
\bigskip
\end{figure}

%\vspace*{1cm}

\vsass
\subsection{The Quark--Hadron Phase Transition}  \label{4.2}

%It is believed that the quarks are free at high temperatures.
%The critical temperature $T_c$ is
%roughly equal to the lowest mass of the hadronic excitations, the pions.
Investigations of the cosmological quark--hadron phase transition started
over a decade ago [\cite{Olive81}; \cite{Suhonen82}]. After these early studies
it was soon realized that the onset of the supposedly first-order
transition required supercooling [\cite{Hogan83}].
%During the subsequent years, several authors
%contributed to our understanding of the transition:
Although much progress was made during the subsequent years
in understanding various features of the transition
[\cite{Witten84}; DeGrand and Kajantie 1984;
\cite{ApplegateHogan85}; Kajantie and Kurki-Suonio 1986;
Fuller, Mathews and Alcock 1988],
%a good reference list can be
%found from the review article by Bonometto and Pantano 1993).
%However,
several questions are still unanswered, partially due
to insufficient knowledge of the properties of thermal
quantum chromodynamics.
(For a more complete reference list see
Bonometto and Pantano [1993].)

Lattice Monte Carlo simulations provide the best tool currently available
for the study of the equation of state in QCD near the transition point.
But even with this method one has not been able to solve
the order of the transition for the
cosmologically relevant case, {\em i.e.}, when the chemical potential
vanishes and there are two light (u,d) and one intermediate-mass (s)
quark species.
What is known is that in the case of pure glue,
corresponding to infinitely heavy quarks,
the transition is first order, and so it is with four light quarks.
(For a review on lattice results see [\cite{Petersson92}].)
Also is known that there is a substantial jump in
the energy density within a temperature interval of less than 10~MeV
around the critical temperature.
However,
for working out the consequences of the phase transition in cosmology
this information is not sufficient as long as one is not
able to distinguish between a first-order, and a second-order
or non-existing transition.
If the transition is not first order,
no supercooling can occur, even if the equation of state gave rise to
a very rapid change in the energy density.
This is due to the extremely slow expansion of the Universe.

Guided by the recent lattice calculations, we use in numerical estimates
the following values for the physical quantities of the quark--hadron phase
transition:
\be
  T_c = 150 \: {\rm MeV} \; , \;\;\;\; L = 2 \, T_c^4 \; , \;\;\;\;
  \sigma = 0.02 \, T_c^3  \; . \la{QHparams}
\ee
Here $T_c$ is the critical temperature, $L$ the latent heat and $\sigma$ the
interface tension.
The true value of the critical temperature lies very probably somewhere between
100 and 250~MeV, and 150~MeV may be a good guess for its value
[\cite{Petersson92}]. For the other two quantities
there is currently no lower limit,
since they vanish if the phase transition is not
first order. The values of $L$ and $\sigma$ given in \eq(\ref{QHparams})
are based on pure glue lattice simulations:
the value of the latent heat is taken from Iwasaki \etal [1992],
and the interface tension has been determined in computer studies
by Grossmann and Laursen [1993], where
the length of the lattice in the imaginary time direction was 2 lattice points,
and by Iwasaki \etal [1993], where it was 4 and 6 lattice points.

The simplest analytical model for the QCD equation of state is that of
the MIT bag model, which is often employed in the cosmological context.
In this model the pressures of quark and hadron phases are given by
\be
  p_q (T) = a_q T^4 - B \; , \;\;\;\;
  p_h (T) = a_h T^4  \; . \la{pQHbag}
\ee
The value of the bag constant $B$ is determined from the
condition of equal pressure at $T_c$: $B \! = \! (a_q \! - \! a_h) T_c^4$.
The bag equations (\ref{pQHbag})
also follow as a special case from the more general equation
of state presented in Subsection~\ref{4.1}.
Approximating $\tilde{B} (T) = (L/4)(1 - T^4/T_c^4)$,
which in the limit of small supercooling is equivalent to \eq(\ref{Deltap}),
and substituting this to the equation of
state~(\ref{pEW}) with the constant $- \! \tilde{B} (0)$ added
in both sides, one recovers the bag equations:
\be
  p_q (T) = a T^4 - \frac{L}{4} \; , \;\;\;\;
  p_h (T) = ( a - \frac{L}{4 T_c^4} ) \, T^4 \; . \la{pQH2}
\ee
%where $a = (a_q \! + \! a_h)/2 + L/8$.

%It would be tempting to think that now we only have to determine the
%pressure coefficients of the bag model by counting the particles
The equation of state of the naive bag model,
which follows from counting the particle species
and utilizing \eq(\ref{pressure}),
gives an approximate upper limit for the latent heat.
Somewhat above the critical temperature $T_c$ the strongly
interacting relativistic particles
are the gluons, and u-- and d--quark.
(Depending on its mass, also the s--quark could be counted.)
For temperatures somewhat below $T_c$ the only strongly interacting particles
that we include within this naive approach are the pions,
the lightest hadrons. The other low-massed and massless particles---the
photon, electron, muon and neutrinos---are present in both phases.
The effective number of relativistic degrees of freedom
in the quark--gluon and in the hadron phase is thus
%, but not too near $T_c$,
given by $g_{*q} \! = \! 51.25$ and $g_{*h} \! = \! 17.25$, respectively.
%The equation of state that follows from plugging these numbers in is here
%called the naive bag model.
%The naive bag model follows from extrapolating the observed high- and
%low-temperature behavior to the critical temperature.

\begin{figure}[t]
%\pspict{5cm}{disk$b:<ignatius.tex.phd>qhepsoft.ps}
\vspace*{5cm}
%Sohvi:Die Hard 40/Janne/Kuvia/PhD/QHepsofT
\caption[a]{Behavior of energy density in QCD.
  Thin curve is for the naive bag equation of state,
  and thick curve for a weaker first-order transition.
  Dotted lines denote the metastable branches.\la{fig:qhepsoft}}
\medskip
\end{figure}

In \fig\ref{fig:qhepsoft}, the energy density of the cosmic fluid
is schematically plotted both
for the naive bag model, and for a more realistic equation of state
consistent with the lattice simulations.
We clearly see how the naive bag model exaggerates the value of latent heat,
$L_{\rm bag} \! = \! 14.9 \, T_c^4  \gg  L$.
The parameter values of the bag model can be corrected to reproduce
a desired latent heat.
However, the corrected bag model does not mimic
well the realistic equation of state over the whole range.
But in the vicinity of $T_c$ it
reproduces the true equation of state with a first-order accuracy in the
pressure and zeroth-order accuracy in the energy density, which is sufficient
for determining the onset of nucleation.
%For that purpose,
In this case the use
of the bag model with improved parameter values is in practice equivalent
to employing directly
the thin-wall approximation discussed in Subsection~\ref{3.4},
except that changing the equation of state also affects the
relation between time and temperature of the Universe.

%During supercooling the value of $g_{*q}$ must be corrected in such a way that
%at $T_c$, the energy density of the quark--gluon phase
%in the corrected bag model is equal to that in the ``real'' equation of state.
%Assuming that the true discontinuity in the energy density is in middle of
%%that
%of the naive bag model, the corrected value we obtain
%is $ \tilde{g}_{*q} \! = \! 36.53$, with latent heat as in
%%\eq(\ref{QHparams}).
%
%However, better results can be achieved within the bag model by rechoosing
%the bag parameters so that the model gives the correct latent heat.
%
%In the thin-wall limit this is sufficient for determining the moment
%of onset of bubble nucleation.
%In practice, this is equivalent to employing the thin-wall formulas from
%Subsection~\ref{3.4} and assigning a new value to $g_{\*q}$. We fix it
%by demanding that

Validity of the thin-wall approximation is violated if the correlation
length associated with the transition is not clearly smaller than
the radius of the critical bubble.
It is not easy to tell what the relevant correlation length is in QCD near
the transition temperature. However, since the transition is just weakly first
order, if first order at all, it is quite possible that the correlation
length is quite large.

In a case where the thin-wall approximation is inapplicable,
one could employ in nucleation calculations the order parameter
model presented in Subsection~\ref{4.1}.
There is a one-to-one correspondence between
the four parameters of the potential (\ref{Veikr}) of that model and
the four physical parameters of the transition
[Paper~I; \cite{Kajantie92}]:
\be
  \left[ T_0, \gamma, \alpha, \lambda \right]  \longleftrightarrow
  \left[ T_c, \sigma, L, l_c \right]  \; , \la{basises}
\ee
where $l_c$ is correlation length. (In the order parameter model with a quartic
potential the correlation lengths in both phases are equal to $l_c$ at
the critical temperature.) Once the values of the physical parameters
are known, also the parameters of the potential are completely fixed.

It is not clear how one should interpret the
order parameter field in QCD because the theory does not have any
classical potential driving the transition.
If the latent heat is not small, the order parameter
represents several degrees of freedom. Then it would seem more natural to
identify the $\phi$--field with a thermodynamical quantity like the energy
density, in the same manner as was done by Csernai and Kapusta~[1992a].

{}From now on we will assume that the nucleation in the cosmological
quark--hadron phase transition took place under conditions which were close
to the thin-wall limit.
For the values of the physical parameters presented in \eq(\ref{QHparams}),
\eq(\ref{Rcrtw}) gives for the critical bubble
%at nucleation
a very large value of the radius,
$R_{\rm cr}(T_f)  =  38/T_c  = 51$~fm.
This shows that the thin-wall approximation
would remain valid even if the correlation length were large.

The main events of the quark--hadron phase transition
are shown in \fig\ref{fig:QH} in the same way as was done
in \fig\ref{fig:EW} for the electroweak case.
Inspecting first the early stages of the phase transition,
we note that in the quark--hadron phase transition
the supercooling was smaller than
in the electroweak transition. Secondly, we may compare the growth time of
shocks with the duration of supercooling
utilizing for example \eqs(\ref{Tftw}) and~(\ref{Rnucltw}):
\be
  \frac{ \Delta t_{\rm sh} }{ \Delta t_{\rm sc} }
  = \frac{ \pi^{1/3} }{ \tilde{A} } \; . \la{tshtsc}
\ee
It is interesting to note that this ratio is completely determined by the
nucleation action, or approximately
by the age of the Universe. In other words, at any
temperature there is the definite relation~(\ref{tshtsc}) between the degree of
supercooling and the nucleation distance
(assuming that the value of the shock velocity is a constant).
This result is in principle valid only in the thin-wall
limit. However, from \fig\ref{fig:EW} one can infer that the prediction
holds rather well also for the electroweak transition
in the example case we considered.

\begin{figure}[p]
%\pspict{13cm}{disk$b:<ignatius.tex.phd>qh.ps}
\vspace*{13cm}
%Sohvi:Die Hard 40/Janne/Kuvia/PhD/QH
\caption[a]{Schematic 1+1 dimensional figure of the phase transition
  from the quark--gluon (grey) to the hadron phase (white).
  This figure was presented originally by Rummukainen [1990];
  the current version is a modified one.
  For clarity only the effect of the dying
  quark droplets is shown in the world lines of the cosmic fluid
  (thin dotted lines), the effect of an expanding hadron bubble
  has been illustrated earlier in \fig\ref{fig:tx}.
  The numbers correspond to the values of physical quantities
%  calculated in the thin-wall approximation
  in \eq(\ref{QHparams}).\la{fig:QH}}
\medskip
\end{figure}

The values derived for
the two time scales $\Delta t_{\rm sc}$ and $\Delta t_{\rm sh}$
may be somewhat erroneous, since
in three dimensions the shocks are weak, especially if the deflagration front
is very slow [\cite{Kurki-Suonio85}]. It seems that the deflagration front
velocity was indeed quite small, probably at least one order of magnitude
smaller than the velocity of the shock front
[\cite{Kajantie92}; Ignatius, Kajantie, Kurki-Suonio and Laine 1993].
The weakness of the shocks could
make the estimates of the nucleation process inaccurate, because
new bubbles could possibly nucleate to a region already touched by a shock.
This inaccuracy can be at least partially cured by using in the calculations
an effective velocity $v$, which is smaller than the true velocity of the
deflagration front.

The main difference between the electroweak and the quark--hadron transition
is that only in the latter transition the Universe was reheated back to
the critical temperature (see \fig\ref{fig:Tt}).
This is due to the much larger
value of the latent heat in the quark--hadron transition.
After the reheating, the phase transition proceeded very slowly, and almost in
thermal equilibrium. The expansion of the Universe did not cause any cooling;
instead, the denser quark--gluon matter was transformed to the more dilute
hadron matter. As is discussed in Paper~I, the duration of this period
can be approximately determined from the relation
\be
  \frac{t_e}{t_c}  - 1 \approx \frac{L}{2\, \varepsilon_q (T_c)} \; . \la{te}
\ee
This approximation holds if the resulting value is clearly smaller than
unity.
%Conservation of the energy of the whole Universe
In the case of this period of slow burning
the expansion rate of the Universe determines
the typical velocities of deflagration fronts, too.
The velocities are
roughly given by $R_{\rm nucl} / R_{\rm hor}$ divided by
%the value from \eq(\ref{te}).
$L/4 \varepsilon_q$.
The numerical value that this gives
for the velocity is of the order of $10^{-4}$.

At the final stages of the phase transition the decaying quark droplets
produced rarefaction waves [\cite{KajantieKurki-Suonio86}]. This led to
the creation of density inhomogeneities, which in principle could
have significantly affected the
nucleosynthesis and could be observed in the present-day Universe.
However, if the parameter values in \eq(\ref{QHparams}) are roughly correct,
the distance scale of density inhomogeneities was too short
%by at least one order of magnitude
for this to happen.
Only a distance scale $R_{\rm nucl}$ of at least one meter at the quark--hadron
transition temperature could have had later an effect on the abundance
of light elements
[Applegate, Hogan and Scherrer 1987; \cite{Kurki-Suonioetal90};
\cite{Mathewsetal90}].
Redshifted to the present Universe,
the length $R_{\rm nucl} \! \approx \! 4 \: {\rm cm}$
is less than the distance from Earth to Sun.
In the cosmic scale this is a very short distance.

If the interface tension were the same as in \eq(\ref{QHparams}) but
the latent heat much smaller, the distance scale $R_{\rm nucl}$,
given in \eq(\ref{Rnucltw}) and
proportional to $\sigma^{3/2} /L$, would correspondingly be much larger.
%Although this possibility cannot presently be ruled out,
%it does not seem probable if
%the order parameter model presented in Subsection~\ref{4.1} can be trusted
%here.
%But the magnitude of supercooling,
%given in \eq(\ref{Tftw}), is almost directly proportional to the
%parameter \mbox{$C \! \simeq \! \sigma ^{3/2} /L \sqrt{T_c}$}
%which scales like \mbox{$C \propto \alpha^{5/2}$}.
%However, since the true
%values of these quantities could be arbitrary small on one hand, and
%cannot be much larger than the values~(\ref{QHparams}) on the other hand,
%the QCD phase transition can at most be weakly first order.
This possibility cannot presently be ruled out, because the values of the
physical quantities $\sigma$ and $L$ are not known.
However, it is tempting to think that the values of these quantities
would not be
arbitrary if the phase transition were very weakly first order; that instead
in this limit they would show some sort of universal behavior
compared with other physical transitions.
By weakly first order transition we mean a transition in which the values of
the thermodynamical quantities $L$, $\sigma$ and $1/l_c$ are in dimensionless
units small when compared with unity.

A phase transition which can be treated analytically is the one
between ordered and disordered phases in the
two-dimensional $q$--state Potts model [see {\em e.g.}\ \cite{Wu82}].
This model is a generalization of the Ising model to $q$ spins, and
the transition is for $q > 4$ first
order, the stronger the larger $q$ is.
In the case of the transition of the two-dimensional Potts model
there are no additional parameters besides $q$.
The values of the latent heat [{\em e.g.}, \cite{Wu82}] and more recently,
the interface tension~[\cite{BorgsJanke92}],
have been calculated analytically.
In the limit where the phase transition becomes weaker and weaker, that is,
when $q$ approaches 4 from above, the interface tension vanishes more
rapidly than the latent heat.
%This conclusion holds even if the assumption of
%complete wetting by Borgs and Janke is not valid, since then their result for
%the interface tension is an upper limit for the true value.
(A similar behavior can be seen in the quartic order parameter model presented
in Subsection~\ref{4.1}, if one believes in the naive way of weakening the
phase transition: by decreasing the value of $\alpha$ and keeping the other
parameters appearing in the potential~(\ref{Veikr}) constant one observes that
\mbox{$\sigma^{3/2}/L \: \propto \: \alpha^{5/2}$}.)
The scaling argument coming from the two-dimensional Potts model
seems to hint that
if a cosmological first-order phase transition is made weaker,
the distance scale between nucleation sites gets smaller---a quite natural
behavior.

It seems that in the cosmological quark--hadron phase transition
the expanding deflagration bubbles were on the borderline between
stability and instability [\cite{Huetetal92}], if one assumes that latent
heat was carried away from the front by hydrodynamic flow and not by
neutrinos. The opposite assumption has also been made
[\cite{FreeseAdams90}; \cite{AdamsFreeseLanger93}], and it easily leads to
instabilities. However, according to Applegate  and Hogan [1985] the relative
importance of neutrinos as heat carriers vanishes
in the limit of small supercooling.

Finally, let us mention two exotic topics related to the cosmological
quark--hadron phase transition. Firstly, Mardor and Svetitsky [1991] made the
observation that in the MIT bag model small hadron bubbles exist already
above~$T_c$.
But this does not imply that the same would be true for the
real QCD. Indeed, employing the quartic order parameter model the normal
behavior is recovered~(Paper~III).
Secondly, in some circumstances
%[check: what circumstances?]
it might be possible
for the quark droplets to survive over the transition [\cite{Witten84}].
If stable, these lumps of strange quark matter could then in principle
be observed in the present-day Universe.

\clearpage

\vsas
\section{Conclusions and Future Prospects}   \label{5}

The goal of this Thesis has been to achieve an insight to the physical
processes that occurred in the Universe during different first-order
phase transitions.
%supposing they were first order.

We have pointed out that especially the onset of
a cosmological phase transition shows a universal behavior
allowing for a general approach.
We have presented in this work some general methods which can be applied to an
analysis of the sequence of events taking place in cosmological first-order
phase transitions.
%As a consequence,
%this sequence of events can be analyzed applying the general methods presented
%in Section~\ref{3}.
%of this introductory review part of the Thesis.

In the case of the cosmological electroweak or quark--hadron phase transition,
the quantitative description has not yet reached a fully satisfactory level.
The reason for this is the lacking knowledge of the correct input
values of the physical parameters.
%related to these transitions.
%The main uncertainty related to
The case with the phase transition suggested in Paper~II is different in the
sense that there the main uncertainty
comes from the basic assumptions of the scenario itself.
%, whereas the values of the physical
%quantities are known more accurately than in the usual transitions.

In the future one will hopefully learn the correct values of the physical
quantities related to the electroweak and quark--hadron phase transitions.
With their improving accuracy
lattice simulations should provide this information
for QCD with physical quark masses, as well as for the electroweak theory.
%For QCD with physical quark masses, lattice simulations form the tool that
%is expected to provide us this information

Combining lattice simulations with analytical calculations
in effective three-dimensional theories, it seems to be possible to derive
for the electroweak theory
a potential incorporating even non-perturbative effects
[\cite{Shaposhnikov93}; \cite{Farakosetal93}].
Such a potential could then be employed for studying the electroweak
phase transition.

In order to understand in detail the dynamics of cosmological phase transitions
one must have a good knowledge of the bubble growth at microscopic level.
At present we are investigating the growth of bubbles using a model in which
there is a friction-like coupling between the order parameter  field and a
cosmic fluid field.  By employing this model one is able to numerically
simulate for instance  the collisions between bubbles. Moreover, the velocity
of a deflagration wall can be determined exactly as a function of the friction
coefficient. So far, we have applied the model in 1+1 dimensions [Ignatius,
Kajantie, Kurki-Suonio and Laine 1993]. In the future, the computations should
be extended  to include spherically symmetric three-dimensional bubbles.
Furthermore, one could try to obtain a good estimate for the value of  the
friction coupling that determines the velocity of the bubble wall.

By combining all these developments, one is led to the conclusion that
it is possible to achieve
an accurate quantitative description of different cosmological phase
transitions in the not too distant future.
It would then also be possible to calculate the value of the baryon asymmetry
of the Universe from the first principles.

\clearpage

\setlength{\baselineskip}{.6cm}

\addtocontents{toc}{ \protect\contentsline {section}{Original Papers}{} }
\addtocontents{toc}{\protect\addvspace{3mm}}

\end{document}